\documentclass[12pt,preprint]{aastex}

\def\gs{\mathrel{\raise0.35ex\hbox{$\scriptstyle >$}\kern-0.6em
\lower0.40ex\hbox{{$\scriptstyle \sim$}}}}
\def\ls{\mathrel{\raise0.35ex\hbox{$\scriptstyle <$}\kern-0.6em
\lower0.40ex\hbox{{$\scriptstyle \sim$}}}}

\shorttitle{Tracing molecular gas mass in extreme environments}
\shortauthors{Zhu, Papadopoulos, Xilouris,  Kuno \& Lisenfeld}

\def \eg           {{e.g.}}

\def \ie           {{i.e.}}

\def \kms          {\hbox{km$\,$s$^{-1}$}}
\def \kkms         {\hbox{K$\,$km$\,$s$^{-1}$}}

\def\approxlt{\lower.2em\hbox{$\buildrel < \over \sim$}}
\def\approxgt{\lower.2em\hbox{$\buildrel > \over \sim$}}

\def \ls           {\hbox{L$_{\odot}$}}
\def \ms           {\hbox{M$_{\odot}$}}           % Solar mass
\def\arcsec{$^{\prime\prime}$}
\def\XdVunit{{\rm\,pc\,{[km\,s$^{-1}$]}$^{-1}$}}
\def\Xunit{{\rm\,cm$^{-2}$\,{[K km\,s$^{-1}$]}$^{-1}$}}

\def\um{{$\mu$m~}}
\def\nh2{{$n({\rm H_2})$}}
\def\cm3{{\,$\rm cm^{-3}$}}
\def\H2{{$H_2$}}
\def\Tk {{\rm $T_{\rm kin}$}}

\begin{document}
\title{Tracing molecular gas mass  in  extreme
 extragalactic environments: an observational study}

\author{Ming Zhu}
\affil{Joint Astronomy Centre/National Research Council Canada, 
660 N. A'ohoku Place, University Park,
Hilo, Hawaii 96720, USA}
\email{m.zhu@jach.hawaii.edu}

\author{Padeli \ P.\ Papadopoulos}
\affil{Argelander Instit\"ut f\"ur Astronomie, Auf dem H\"ugel 71, 53121 Bonn,
Germany}
\email{padeli@astro.uni-bonn.de}

\author{Emmanuel \ M.\ Xilouris}
\affil{Institute of Astronomy and Astrophysics, National Observatory of Athens, 
P. Penteli, 15236 Athens, Greece}
\email{xilouris@astro.noa.gr}

\author{Nario Kuno}
\affil{Nobeyama Radio Observatory, Minamimaki, Minamisaku, Nagano 384-1305, Japan}
\email{kuno@nro.nao.ac.jp}

\and
\author{Ute Lisenfeld}
\affil{Dept. F\'\i sica Te\'orica y del Cosmos, Universidad de Granada, Spain}
\email{ute@ugr.es}

\begin{abstract}
  
  We present a new observational study of the $^{12}$CO(1--0) line
  emission as an H$_2$ gas mass tracer under extreme conditions in
  extragalactic environments.  Our approach is to study the full
  neutral interstellar medium (H$_2$, \ion{H}{1} and dust) of two
  galaxies whose {\it bulk} interstellar medium (ISM) resides in
  environments that mark (and bracket) the excitation extremes of the
  ISM conditions found in infrared luminous galaxies, the starburst
  NGC\,3310 and the quiescent spiral NGC\,157. Our study maintains a
  robust statistical notion of the so-called
  $X=N({\rm{H_2}})/I_{\rm{CO}}$ factor (i.e.  a large ensemble of
  clouds is involved) while exploring its dependency on the very
  different average ISM conditions prevailing within these two
  systems.  These are constrained by fully-sampled $^{12}$CO(3--2) and
  $^{12}$CO(1--0) observations, at a matched beam resolution of Half
  Power Beam Width $\sim 15''$, obtained with the James Clerk Maxwell
  Telescope (JCMT) on Mauna Kea (Hawaii) and the 45-m telescope of the
  Nobeyama Radio Observatory (NRO) in Japan, combined with sensitive
  $850\,\mu$m and $\rm 450\,\mu$m dust emission and \ion{H}{1}
  interferometric images which allow a complete view of all the
  neutral ISM components.  Complementary $^{12}$CO(2--1) observations
  were obtained with JCMT towards the center of the two galaxies.  We
  found an $X$ factor varying by a factor of 5 within the spiral
  galaxy NGC\,157 and about 2 times lower than the Galactic value in
  NGC\,3310. In addition, the dust emission spectrum in NGC\,3310
  shows a pronounced submm ``excess''.  We tried to fit this excess by
  a cold dust component but very low temperatures were required
  ($T_{\rm C}\sim 5 - 11$ K) with a correspondingly low gas-to-dust
  mass ratio of $\sim 5 - 43$.  We furthermore show that it is not
  possible to maintain the large quantities of dust required at these
  low temperatures in this starburst galaxy. Instead, we conclude that
  the dust properties need to be different from Galactic dust in order
  to fit the submm ``excess''.  We show that the dust SED can be fitted
  by an enhanced abundance of Very Small Grains (VSGs) and discuss
  different alternatives.

\end{abstract}

\section{Introduction}

The use of the $^{12}$CO(1--0) line emission as a tracer of H$_2$ gas
has become an established technique since its the first detection
(Wilson, Jefferts \& Penzias 1970), with several large surveys using
it to obtain images of its distribution in galaxies across the Hubble
sequence (e.g.  Young \& Scoville 1991; Helfer et al.  2003; Leroy et al.
2005) and examine its relation to star formation (e.g.  Wong \& Blitz
2002). Several theoretical studies have been published since in order
to outline the dependence of the so called $ X$ factor converting the
luminosity of this optically thick tracer to H$_2$ mass under the
ambient ISM conditions (e.g.  Dickman, Snell, \& Schloerb 1986;
Wolfire, Hollenbach, \& Tielens 1993; Bryant \& Scoville 1996). These
have now converged towards a view of a fairly robust $X$ factor
provided that: a) the $^{12}$CO(1--0) line is well-excited, b) a large
ensemble of CO-bright molecular clouds is involved (so the statistical
notion of $ X$ remains valid), c) the molecular gas observed is
reducible to a set of self-gravitating ``units'' (whose linewidths
reflect solely their gas mass), and d) for a given far-ultraviolet
(FUV) interstellar radiation field the molecular gas is sufficiently
metal-rich to ensure that CO remains abundant throughout the volume of
a typical molecular cloud ($Z\ga 1$ for $G_{\circ}=1$ in Habbing
units).  The latter restriction stems from the fact that unlike H$_2$
the CO molecule, with its much lower relative abundance ($\rm
[CO/H_2]\sim 10^{-4}$), cannot self-shield from the dissociating FUV
radiation and its relative extent in molecular clouds is regulated by
the absorption of that radiation by dust (Pak et al. 1998).

The strong dependence of the $X$ factor on metallicity has been
expected from theoretical studies (Maloney \& Black 1988; Bolatto,
Jackson, \& Ingalls 1999), and confirmed by extensive observations of
the Magellanic Clouds (Israel 1997, 1999) and the Magellanic irregular
IC 10 (Madden et al. 1997).  In the high-excitation ISM environments
found in the Ultra Luminous Infrared Galaxies (ULIRGs) an unbound
molecular gas phase (a result of their violent, merger-induced,
kinematic environments, e.g.  Aalto et al.  1995) is thought as
responsible for a systematic overestimate of the H$_2$ gas mass when a
Galactic $X$ factor $X_{\rm{G}}$ is used (Solomon et al.  1997; Downes
\& Solomon 1998). Towards the other excitation extreme, a cold and/or
diffuse gas phase with very low $^{12}$CO(1--0) brightness has been
discovered in molecular clouds in the nearby quiescent galaxy M 31
(Allen \& Lequeux 1993; Allen et al.  1995; Loinard, Allen, \& Lequeux
1995), as well as in the outer Galaxy, leading even to suggestions
that cold molecular gas constitutes an important component of the dark
matter in spirals (Lequeux, Allen \& Guilloteau 1993).

Unlike the relatively well-explored metallicity dependence of the $X$
factor, the effects of a highly excited, possibly non
self-gravitating, molecular gas phase in starburst environments, and a
low-excitation one in quiescent ones harboring low star formation
rates, have not been widely studied.  For ULIRGs such efforts have
been hindered by their compact CO emission regions, the presence of a
``standard'' ensemble of star-forming molecular clouds mixed with the
highly-excited phase, and the current inability to conduct multi-line
CO observations at high and beam-matched resolutions.  The latter
could disentangle the two gas phases present in those systems, and
such observations of ULIRGs will be important in the future.  In low
excitation environments, identified by the presence of cold dust
and/or low $^{12}$CO(2--1)/$^{12}$CO(1--0) line ratios ($r_{21}$),
lack of data constraining the gas excitation allows a degeneracy
between a diffuse and a cold gas phase, which translates to large
uncertainties in the estimated H$_2$ gas mass (e.g.  Israel, Tilanus,
\& Baas 1998).  Moreover, while it can be argued that very cold
molecular clouds do exist in a few places in galaxies, it is far from
certain that they can be typical for a significant fraction of their
total molecular gas mass, even in quiescent environments (e.g.  Wilson
\& Mauersberger 1994; Sodroski et al.~1994).

From the largest CO line ratio survey available (Braine \& Combes
1992, BC92 hereafter) we chose two infrared luminous galaxies,
NGC\,3310 (a starburst) and NGC\, 157 (a quiescent spiral), whose
physical conditions of the {\it bulk} of their molecular gas and dust
mass best represent the excitation extremes within the IRAS
Intermediate Luminosity Sample [$L_{IR}=(10^{10}-10^{11})\,L_{\odot
}$] (Tinney et al.  1990). Both have similar far-infrared (FIR)
luminosities ($\sim 10^{10}\,L_{\odot}$) but very different
$S_{100}/S_{60}$ ``colors'' of $\sim 1.28$ (NGC\,3310) and $\sim 2.42$
(NGC\,157) while their reported $^{12}$CO(2--1)/$^{12}$CO(1--0) line
ratios mirror this with $r_{21}$(NGC\,3310)$\sim 2.6$ (the highest
reported such ratio, but see \S 3.1.1), and $r_{21}$(NGC\,157)$\sim
0.4$ (typical for cold quiescent molecular clouds in the outer
Galaxy).  We make a comparison study of their ISM properties using the
$^{12}$CO(3--2) and $^{12}$CO(1--0) lines, along with $\rm 850\,\mu m$
and $\rm 450\,\mu m$ dust continuum maps and sensitive \ion{H}{1}
images obtained from the literature.  Our main goal is to investigate
excitation effects on the H$_2$ gas mass estimates in the
extragalactic domain.

The optically thin dust continuum emission at FIR and
submillimetre wavelengths can be used to measure the dust column
density, and thus provide an alternative way to trace the gas mass in
most ISM environments where gas and dust are well mixed. Such an approach
is independent of gas excitation and can potentially detect very cold
gas in regions where CO is not excited (Bernard et al. 2008), but its use in
extreme extragalactic environment is hindered by a large uncertainty
in the gas-to-dust (GD) mass ratio. We compute the GD mass ratios in our
sample galaxies and explore whether we can assume a Galactic GD ratio
to estimate the gas mass from dust continuum emission.

In \S 2 we describe our observations and data reduction procedures. 
In \S 3 we present our results from the molecular line analysis
as well as the dust SED modeling, while in \S 4 we discuss the extreme
ISM environments that the two galaxies under study represent. Finally,
in \S 5 we summarize our conclusions. Appendix A shows how the interstellar
radiation field for the two galaxies was calculated.

Throughout the paper we assume a Hubble constant of $H_0=50 $ km
s$^{-1}$ Mpc$^{-1}$.  The distances in our calculations are assumed to
be 21.3 Mpc for NGC\,3310 and 35.0 Mpc for NGC\,157 (Stevens et al.,
2005).

%%%%%%%%%%%%%%%%%%%%%%%%%%%%%%%%%%%%%%%%%%%%
%\section{Observations}\label{observations}
%%%%%%%%%%%%%%%%%%%%%%%%%%%%%%%%%%%%%%%%%%%%

\section{Observations and data reduction}

We mapped the $^{12}$CO(1--0) emission of both galaxies with the
Nobeyama 45-m telescope\footnote{The Nobeyama Radio Observatory (NRO)
is a division of the National Astronomical Observatory of Japan (NAOJ)
under the National Institutes of Natural Sciences (NINS).}  and the
$^{12}$CO(3--2) emission with the James Clerk Maxwell Telescope
(JCMT)\footnote{The JCMT is operated by the Joint Astronomy Center in
Hilo, Hawaii, on behalf of the parent organizations Science and
Technology Facilities Council in the UK, the National Research Council
of Canada, and the Netherlands Organization for Scientific Research.}.
Given the importance of accurate measurements of the
$^{12}$CO(3--2)/$^{12}$CO(1--0) line ratio as a function of position
in our present study, we devoted a large amount of our observing time
at both telescopes towards pointing checks and observations of
spectral line standards.  We also used the two telescopes to observe
$^{12}$CO(3--2) and $^{12}$CO(1--0) towards a common position in
Orion-KL.  The physical conditions of the H$_2$ gas towards this
region are constrained by a multitude of molecular line observations
reaching as high as $^{13}$CO(5--4), with inferred densities of $\rm
n(H_2)\ga 10^4\, cm^{-3}$ and $ T_{\rm{kin}}\sim 45$ K (Plume et al.
2000).  For such conditions the intrinsic
$^{12}$CO(3--2)/$^{12}$CO(1--0) brightness temperature ratio expected
is $R_{31}\sim 1$ and thus for the similar beams of the JCMT and the
Nobeyama 45-m telescope: $ T_{\rm mb}(3-2)/T_{\rm mb}(1-0)\sim
R_{31}\sim 1$.  Our calibration observations found very good agreement
for associated spectra from both telescopes, with an averaged
integrated intensity ratio of 0.93.  This ensures that there are no
serious calibration offsets between the two telescopes.

\subsection{CO observations}

\subsubsection{$^{12}$CO(3--2) and $^{12}$CO(2--1) observations at the JCMT}

We used the JCMT during the periods of July 1999, January 2000 and
February 2001 to conduct $^{12}$CO(3--2) imaging of the starburst
galaxy NGC\,3310 and then again in June 2002, and May, June and July
of 2004 for similar observations of the quiescent spiral NGC\,157.  In
all cases we employed receiver B3 in a dual channel mode, tuned to
Single Side Band (SSB) to observe the $^{12}$CO(3--2) line at
$345.796$ GHz and used the Digital Autocorrelation Spectrometer (DAS)
with a bandwidth of 920 MHz ($\rm \sim 795\, km\, s^{-1}$). Both maps
were acquired with a $\Delta \theta = 1/2\, \theta _{\rm HPBW}=7''$
sampling, which resulted in the acquisition of spectra in 47 positions
for NGC\,3310, and 194 positions in NGC\,157, covering most of the
dust emission detected in our submillimeter (submm) continuum maps of
these galaxies.  We employed rapid beam-switching at a frequency of 1
Hz and chop throws of $\sim 120''-180''$ (in Az).  For the NGC\,3310
observations typical system temperatures where $T_{\rm sys}\sim (400 -
600)$ K with $\Delta t_{\rm int}\sim 20$ min of integration time per
point (effective time for dual channel operation), while for NGC\,157
these figures were $T_{\rm sys}\sim (300 - 400)$ K, and $\Delta
t_{int}\sim (10-20)$ min.  The pointing was monitored every (45-60)
min and from the pointing residuals we estimated an rms pointing
uncertainty of $\delta \theta _{\rm rms}\sim 2''-3''$.

A position with strong $^{12}$CO(3--2) emission was repeatedly
observed within each galaxy during our imaging observations to verify
proper pointing and consistent line calibration (i.e.  by checking
line profiles and intensities).  These observations along with
numerous ones of spectral line standards yielded a spectral line
calibration uncertainty of $\sim 10\%$. The latter is the 2nd term
contributing to the total uncertainty of line intensities given by

\begin{equation}
  \left(\frac{\delta   T}{T}\right)_{\rm rms}=\left[  \left(\frac{\delta
         T}{  T}\right)^2 _{\rm ther}+ \left(\frac{\delta 
        T}{ T}\right)^2  _{\rm   cal} +  \left(\frac{\delta   
        \eta _{\rm mb}}{ \eta _{\rm mb}}\right)^2  \right]^{1/2},
\end{equation}

\noindent
where $T$ is the main beam brightness temperature of the line,
averaged over a total number of channels $ N_{\rm ch}$, and with a
baseline defined over a total number of channels $ N_{\rm bas}$.  The
second term describes the uncertainties inherent in the calibration of
mm/submm lines (e.g.  temperature uncertainties of cold/warm loads
used in the 3-point calibration) and the 3rd term is the uncertainty
associated with the adopted value of the main beam efficiency $\eta
_{\rm mb}$.  The contribution of the pure thermal noise is given by

\begin{equation}
\left(\frac{\delta T}{T}\right)_{\rm ther}=\frac{\delta T_{\rm ch}}{T}\left(\frac{N_{\rm bas} + 
N_{\rm ch}}{N_{\rm bas} N_{\rm ch}}\right)^{1/2},
\end{equation}

\noindent
where $\delta T_{\rm ch}$ is the thermal rms noise across the
line-free channels of the spectral line.  Repeated observations of
Mars and Uranus yielded beam efficiency estimates at 345~GHz of $\eta
_{\rm mb}=0.62$ for the NGC\,3310, and $\eta _{\rm mb}=0.65$ for the
NGC\,157 observations.  Inspection of the JCMT
database\footnote{http://www.jach.hawaii.edu/JCMT/spectral\_line/Standards/eff\_web.html}
demonstrates a $\sim 10\% $ uncertainty for the $\eta _{\rm mb}$
values.  The latter enters the total rms uncertainty (equation 1) of
the spectral line strengths in our maps (and thus that of the
$^{12}$CO(3--2)/$^{12}$CO(1--0) ($r_{31}$) line ratios), but is not
expected to contribute to the expected rms line strength (and thus
line ratio) variations {\it within} maps.

We also used receiver A3 tuned to Double Side Band (DSB) to observe
the $^{12}$CO(2--1) line (230.538~GHz) towards the center of both
galaxies. The DAS was used at a 1.8 GHz bandwidth ($\sim 2340$ km
s$^{-1}$).  Observations of Mars, Uranus and Jupiter yielded beam
efficiencies of $\eta _{\rm mb}=0.67$ for NGC\,3310 and $\eta _{\rm
mb}=0.71$ for NGC\,157, for a beam of $\theta _{\rm HPBW}\sim 21''$,
values that we adopt for our conversion of the $T^*_{\rm A}$ to the
$T_{\rm mb}$ scale at 230 GHz.

Finally all the spectra per position were inspected individually and we
co-added only  those found to be  consistent within the  S/N and after
removal of  bad channels on  a few occasions.  Linear  baselines were
subtracted and the $^{12}$CO(3--2) spectra were binned to a resolution of 25
MHz ($\sim 21.6$ km s$^{-1}$), divided by the appropriate $\eta
_{\rm mb}$ and then added to the grid maps of the galaxies (Figures~1 and 2).
The maps were convolved to a final resolution of 15$''$ (as
to exactly match that of the 45-m) and then inserted into AIPS for
further processing and combination with the $^{12}$CO(1--0) cubes.

\subsubsection{$^{12}$CO(1--0) observations at the Nobeyama 45-m telescope}

The observations with the 45-m telescope of the National Radio
Observatory (NRO) of Japan were made in December 2000 and December
2004.  We used the Superconductor-Insulator-Superconductor (SIS)
receiver S100 as the frontend and an Acousto-Optic Spectrometer (AOS),
with a frequency resolution of 250 kHz and the total bandwidth of 250
MHz ($\sim 650$ km s$^{-1}$).  The intensity calibration was performed
by the two-point chopper-wheel method, yielding the total $T_{\rm
sys}$, which typically was $T_{\rm sys}\sim (500-700)$ K.  The
telescope pointing was checked every 30\,min by observing SiO maser
emission of late type stars at 43~GHz.  The rms pointing uncertainty
was found to be $\delta \theta _{\rm rms}\sim 4''$.  We used a
position-switching mode with an integration time of 20 sec for on- and
off-source.  We measured the main beam efficiency using planets and
found it to be $\eta _{\rm mb}\sim 0.4 $ with an uncertainty of $\sim
15\%$.  The half-power beam width was estimated from observations of
3C 279 and is $ \theta _{\rm HPBW}\sim 15''$.  We finally obtained
several spectra of $^{12}$CO(1--0) of Orion-KL in order to obtain the
line calibration intensity as well as check the relative calibration
of the Nobeyama 45-m telescope with respect to the JCMT.
The integration time was 10 min for most points, thus the rms is
higher in the $^{12}$CO(1--0) spectra, and the baseline is also worse
than that of the $^{12}$CO(3--2) due to position switching, especially
for the weak $^{12}$CO(1--0) lines.

\subsection{Dust continuum submm observations with SCUBA}

Our SCUBA observations of NGC\,3310 and NGC\,157 were carried out on
the nights 2000 January 24 and 25, with grade 1 and 2 weather
respectively.  The SCUBA submm bolometer array (Holland et al.  1999)
is an array of 37 bolometers at 850\,$\mu$m and 91 at 450\,$\mu$m
covering a region of sky of $\sim$ 2.3\,arcmin in diameter. In order
to provide fully sampled images, the secondary mirror moves in a
64-step jiggle pattern, with the integration time lasting 1\,s at each
position.  After the 16 steps of jiggle pattern, the telescope nods in
order to allow for slowly varying sky gradients.
 
Additionally to the observations obtained by us, we used SCUBA data
from the JCMT archive (see \S 3.3.1).  The data were reduced using the
SURF software package (Jenness et al. 1998).  We made corrections for
the atmospheric absorption using the opacities derived from skydip
measurements taken regularly every night. Noisy bolometers were
flagged and large spikes were removed. During each night, several
calibration sources were mapped in order to flux calibrate our images.
Finally, the individual jiggle maps were co-added, after the proper
reduction was made, and a final map was created for each galaxy. From
the 450\,$\mu$m observations described above, only the ones taken in
grade 1 weather conditions had sufficient signal-to-noise ratio and
were used in our analysis. Flux calibration of the final maps, in
units of Jy beam$^{-1}$, as well as the corresponding beam profiles
were obtained from beam maps of CRL\,618 observed with the same chop
throw as our galaxy maps. The calibration information is summarized in
Table 1.

%\newpage

\section{Results}

\subsection{The CO data}

In Tables 2 and 3 we tabulate the integrated intensity of
$^{12}$CO(1--0), $^{12}$CO(3--2) ($I_{10}$, $I_{32}$) and the
$r_{31}=I_{32}/I_{10}$ ratios for the different positions within
NGC\,3310 and NGC\,157 respectively. The uncertainty in $I_{10}$,
$I_{32}$ is calculated according to equation (1) and the uncertainty
for $r_{31}$ was combined quadratically from that of $I_{10}$ and
$I_{32}$.  In most cases $\sigma_{10}$ is higher than $\sigma_{32}$
for the reasons described in \S 2.1.2, hence the uncertainty in
$r_{31}$ is mostly from $\sigma_{10}$.  A pointing error could cause
extra uncertainty in $ r_{31}$ when the two telescopes were not
observing the same position. The expected error is on average 10-20\%
for a pointing error of 4$''$ at $^{12}$CO(1--0) and 3$''$ at
$^{12}$CO(3--2), but could be in principle as high as $\sim$ 50\% if
both telescopes are mis-pointed in different directions by a full grid
cell size of 7$''$.

In Figures 1 and 2 we present the $^{12}$CO(3--2) (red) and
$^{12}$CO(1--0) (green) profiles overlaid on a grey scale SCUBA image
of the dust emission at 850 $\mu$m of NGC\,3310 and NGC\,157
respectively.  In all positions where high S/N $^{12}$CO(3--2) and
$^{12}$CO(1--0) spectra exist there is good agreement in their line
profiles, demonstrating a good relative pointing on the grid positions
by the two telescopes.

\subsubsection{NGC\,3310: a high excitation case}

For NGC\,3310, $^{12}$CO(1--0) and $^{12}$CO(3--2) emission is
detected in 33 positions covering a $42'' \times 42''$ region,
corresponding to 4.3 kpc $\times$ 4.3 kpc in physical scale (see
Figure 1 and Table 2). The $r_{31}$ ratio is slightly higher than
unity in most positions, and slightly enhanced mostly in the southwest
in comparison to the northeast of the nucleus (though only marginal
considering the S/N).  The area average is $\langle r_{31}\rangle
$=1.17, while the range of values found within the inner region of
NGC\,3310 is $r_{31}=0.9 - 1.4$.  This range is still consistent with
a constant ratio of $\langle r_{31}\rangle $ =$1.15 \pm 0.25$ (the
average value of the extreme $r_{31}$ values mentioned above) given
the $\sim 20\%$ noise variations of this ratio expected within the
map.  These are very high values, suggesting a highly excited molecular
gas phase throughout NGC\,3310.  For comparison, starburst nuclei show
an average ratio of $ r_{31}=0.64$ (Devereux et al. 1994; Yao et
al. 2003; Narayanan et al.  2005).

In the survey conducted by Braine and Combes (1992) NGC\,3310 had the
highest $r_{21}$ reported with $r_{21}$ = 2.6. This value is high even
when compared to the highest $r_{21}$ values found within individual
molecular clouds (molecular line ratios of individual regions within
molecular clouds span a much larger range than such ratios over large
molecular cloud ensemble averages encompassed by most extragalactic
observations).  Interestingly $r_{21}\ga 2$ ratios have only been
observed in the N83/N84 region of the Small Magellanic Cloud (Bolatto
et al.  2003), which is part of an expanding molecular shell centered
on a supernova remnant (SNR; Haberl et al 2000).  For much larger
scales such high ratios have been reported also for the nearby
archetypical starburst galaxy M 82 (Loiseau et al.  1990), but were
not confirmed by subsequent observations (Mao et al.  2000 and
references therein). This underscores the need for beam-matched
observations (that become progressively more difficult with the higher
J-separation of rotational transitions) and good cross-calibration of
mm/submm telescopes for accurate line ratio measurements.

Our own $^{12}$CO(2--1) measurements (Figure 3a) yield
velocity-integrated line intensity $I=\int T_{\rm mb} dV$ of
$I_{21}=9.8\pm 1.6$ K km s$^{-1}$ for NGC\,3310, in excellent
agreement with the value $I_{21}= 9.5$\, \kkms\ reported by BC92.
However for $^{12}$CO(1--0) at the (0, 0) position, after convolving
to a 21$''$ resolution we find $I_{10} =6.62 \pm 2.00$\, \kkms, which
is 1.8 times higher than that reported by BC92.  As a result, $r_{21}$
is $1.48 \pm 0.25$ at the central 21$''$ region, which is
significantly lower than the value reported in BC92, but still
indicative of the highly excited molecular gas that seems to pervade
this galaxy.  Convolving the $^{12}$CO(3--2) map to a 21$''$
resolution yields $I_{32}=9.8\pm 1.8$ and $ r_{31}=1.48$ for the
(0$''$,0$''$) position.  The extreme ISM excitation conditions in
NGC\,3310 revealed by the $^{12}$CO(1--0), $^{12}$CO(2--1), and
$^{12}$CO(3--2) relative line strengths are further corroborated by
the warm IRAS ``color'' ratio of $S_{100}/S_{60}\sim 1.28$ (compared,
e.g. with the value of $\sim 2.42$ for the quiescent spiral NGC\,157).

\subsubsection{NGC\,157: variations in the gas excitation}

For NGC\,157, $^{12}$CO(1--0) and $^{12}$CO(3--2) emission is detected
in 64 positions within a $28'' \times 126''$ region, corresponding to
4.7 kpc $\times$ 21.4 kpc in physical scale (see Figure 2 and Table
3). The $r_{31}$ ratios are slightly higher than unity at the nuclear
region but drop to less than 0.5 in the two spiral arms, in a
remarkably different fashion from the $r_{31}$ distribution seen in
NGC\,3310.  Indeed, quite unlike the starburst NGC\,3310 whose
galaxy-wide star-forming activity leaves no discernible gas excitation
gradient, NGC\,157 is quiescent, with little star formation activity
but with a marked change of ISM physical conditions.  Maps of
$^{12}$CO(3--2) line emission in 12 nearby galaxies studied by Dumke
et al. (2001) have shown such decreasing $r_{31}$ ratios outward from
the centers of many galaxies but none are as large as those measured
in NGC\,157.

High $r_{31}$ ratios are found near the nucleus at offset
($0''$,$0''$) and ($0''$, $7''$), coincident with high SCUBA
$S_{450}/S_{850}$ flux ratios, suggesting a high dust and gas
temperature at this position. Low values of $r_{31}\sim 0.2$ on the
other hand are found in the northeast (NE) and southwest (SW) disk,
near offset ($7''$, $49''$) and (-7$''$, -42$''$) respectively, which
host strong 850~\um emission with a low $S_{450}/S_{850}$ flux ratio
and thus low dust temperature (see \S 3.3.3).

For the central $23''$ of this galaxy Braine et al.  (1993) report
  $I_{10}=23\pm 0.9$\,\kkms.  However, after convolving our
  $^{12}$CO(1--0) data to 23$''$, we find $I_{10}=10.7\pm 1.5$\,\kkms,
  which is $\sim 50\%$ of their value.  We note that our pointing
  center was offset from that of the IRAM observations by ($-14''$,
  $-4''$), and Braine et al. (1993) also put a note on the pointing
  uncertainties in the $^{12}$CO(1--0) and $^{12}$CO(2--1) data.
  Since the $^{12}$CO(1--0) emission is much stronger in the SW disk,
  a strong CO line would show up if the telescope was miss pointed
  toward the SW disk, say at offset (-14,$''$ -7$''$).  Comparing with
  our $^{12}$CO(3--2) and $^{12}$CO(1--0) profiles, the IRAM
  $^{12}$CO(1--0) profile published by Braine et al. (1993) is
  consistent with the profiles in the SW of the nucleus, with the peak
  shifted to the red side and a broad wing on the blue side.  Thus it
  is possible that the IRAM measurement was actually corresponding to
  the SW disk rather than the central nuclear region.  This may be the
  reason for the unusually low $ r_{21}\sim 0.43$ reported by BC92.

Our JCMT $^{12}$CO(2--1) measurements (Figure 3b) yield $I_{21}=
9.7\pm1.8$ K km s$^{-1}$ for NGC\,157 at offset ($0''$,$0''$), with a
resolution of 21$''$.  Convolving the $^{12}$CO(1--0) data to 21$''$
resolution we get $I_{10} = 9.7-9.8$ K km s$^{-1}$ and therefore
$r_{21}$=1.0 in the central 21$''$ region, which is consistent with
the high excitation suggested by the $r_{31}$ ratio towards the same
position.

\subsection{The physical conditions of the molecular gas: a study of extremes}

We employ a large velocity gradient (LVG) model (\eg, Goldreich \&
 Kwan 1974) to constrain the physical parameters and eventually the
 CO-to-H$_2$ conversion factor $X$ in the extreme ISM environments
 marked by the two galaxies. This model assumes that the systematic
 motions rather than the local thermal velocities dominate the
 observed linewidths of the molecular clouds. We use this model to fit
 the observed line ratios for different combinations of (\Tk,
  $n({\rm H_2})$, $\Lambda $), where \Tk is the kinetic temperature and 
 $\Lambda =Z_{\rm CO}/(dV/dr)$, with $Z_{CO}$=[$ ^{12}$CO/H$_2$] being the
 fractional abundance of $^{12}$CO with respect to H$_2$ and $dV/dr$ the
 velocity gradient. The optimum set of parameters is determined by
 minimizing $\chi ^2$.

Table 4 lists the input $r_{21}$ and $r_{31}$ ratios used for our LVG
analysis, and the details of our method are outlined in Zhu et al.
(2003; 2007).  The $X$ factor can be directly derived from the LVG
parameters using the formula
\begin{equation}
X= \frac{n({\rm H_2}) \Lambda }{ Z_{\rm CO} T_{\rm rad}},
\end{equation}

\noindent
where $T_{\rm rad}$ is the radiation temperature for 
the $^{12}$CO(1--0) line transition.
 
In order to reduce the well-known degeneracies in the deduced properties
of molecular gas from radiative transfer modeling of a few line
ratios of the (usually) optically thick $^{12}$CO lines we use the
constraint of $ T_{\rm kin}\geq T_{\rm dust}$ (where $T_{\rm dust}$ is
derived from FIR and submm data in \S 3.3).  This inequality is driven
by FUV-heating of gas and dust (usually not thermally coupled unless
$n({\rm H_2})>10^4$ cm$^{-3}$), the turbulent heating of the gas (but
not the dust), and the much more efficient dust cooling (via continuum
emission) with respect to that of the gas (through spectral line
emission).  As shown in \S 3.3, $T_{\rm dust}$ is 32\,K for NGC\,3310
and 23--28\,K for NGC\,157 nuclear region. Hence we estimate $T_{\rm
kin}$ = 35--50 K for NGC\,3310, and 25--40 K for the central region of
NGC\,157.

A further constraint can be set by adopting only the LVG solutions
  that correspond to gas motions that yield a restricted range of
  $K_{\rm vir}$ values where

\begin{equation}
K_{\rm vir} = \frac{(dV/dr)_{\rm obs}}{(dV/dr)_{\rm vir}} 
\sim 1.54 \frac{Z_{\rm CO}}{\sqrt{\alpha} \Lambda} \left[\frac{n({\rm H_2})}{10^3 {\rm cm}^{-3}}\right]^{-1/2}
\end{equation}

\noindent
must be  $K_{\rm vir}\geq 1 $,  with $K_{\rm vir}=1$  corresponding to the
virialized motions  of fully  self-gravitating gas ``cells''  (Papadopoulos \&
Seaquist 1999) ($\alpha \sim 1 - 2.5$,  depending on the assumed cloud density
profile).  From the  modeling of  molecular  line emission  from quiescent  to
star-forming environments (and  using several $^{12}$CO and $^{13}$CO lines) it was
found that $K_{\rm vir}\sim 1-15$ (e.g.  Papadopoulos \& Seaquist 1999), 
which we adopt here as the permitted range of values.

\subsubsection{NGC\,3310 }

The high $r_{21}$ and $r_{31}$ ratio in this galaxy can only be fitted
with a gas phase that corresponds to an optically thin $^{12}$CO(1--0)
line. Tellingly no fits are possible for $T_{\rm kin}$ $< $ 30 K and n
$<$ 3000 \cm3 under the constraint $K_{\rm vir} \leq 15$.  Similar
constraints have also been reported by Yao et al. (2003) when modeling
equally high $r_{31}$ ratios found for galaxies in their sample.

The best fits are found in the range of $T_{\rm kin}$ $=40-60$ K,
\nh2$ = 5000 - 8000$ \cm3, $\Lambda = 1.0- 2.7 \times 10^{-6} $
\XdVunit, which can fit the $ r_{21}$ and $r_{31}$ ratios within the
observational uncertainties ($\chi^2 < 2$).  For this range the $X$
factor is $1.3 - 2.6 \times 10^{19} /(\frac {Z_{\rm CO}}{ 10^{-4}})$
\Xunit.  If we further require $K_{\rm vir}$ to be minimum, i.e.  a
velocity gradient not much higher than its virial value, the best fit
would then be n=7800 \cm3, \Tk=50 K and $\Lambda = 2.7 \times 10^{-6}$
\XdVunit, with $X= 2.6 \times 10^{19}/( \frac {Z_{\rm CO}}{10^{-4}}) $
\Xunit (Table 5).

A major uncertainty is associated with the adopted CO abundance
$Z_{\rm CO}$, usually in the range $ 0.5 - 2.7 \times 10^{-4} $
(Blake et al.  1987; Hartquist et al.  1998) in star-forming clouds
of our Galaxy.  In the case of NGC\,3310 Pastoriza et al. (1993) found
metal abundances 2.5 to 5 times lower than Solar in its circum-nuclear
\ion{H}{2} regions, and thus the CO abundance could be a factor of 5
lower than the average value of our Galaxy ($10^{-4}$).  For $Z_{\rm
CO}= 2 \times 10^{-5}$, the $X$ factor would be $1.3 \times 10^{20} $
\Xunit (Table 5), which is a factor of 1.5 times lower than the
Galactic value $X_{\rm G}= 2.0 \times 10^{20} $ \Xunit (e.g. Strong \&
Mattox 1996), with the excitation bias partial offset by the
metallicity effect.

In all the above LVG solutions, the opacity at $^{12}$CO(1--0) is
$\tau _{10}<1$, quite unlike in typical Giant Molecular Clouds (GMC)
where this line is optically thick ($\tau _{10}\geq 5$).
Interestingly, only molecular gas in the vicinity of an SNR (and
NGC\,3310 is expected to have lots of them) presents high CO ratios
(Bolatto et al.  2003) and optically thin CO emission.  Even in
extreme starbursts such as (Ultra)Luminous Infrared Galaxies (LIRGs
and ULIRGs) as Arp 220 and NGC\,6240, $r_{21}$ and $r_{31}$ ratios are
all $\la 1$ (Greve et al.  2009).  Thus the molecular gas in NGC\,3310
seems to be indeed defining the uppermost envelope of average
excitation conditions known for IR luminous galaxies in the local
Universe.

%------------------
\subsubsection{ NGC\,157 }
%------------------

The ratios of integrated line intensities for NGC\,157 span a wide
range, but still within that seen in other galaxies (e.g. Yao et al.
2003), making their LVG fitting more straightforward.  At the nuclear
region conditions appear to be similar to those of NGC\,3310 with \Tk
$=30-40$ K, \nh2$ = 2000-5000$ \cm3, and $\Lambda = 2.7- 7 \times
10^{-6} $ \XdVunit (Table 5). The range of kinetic temperature is that
of the dust since, with limited heating sources in the central region
of this galaxy (e.g.  not much star formation or AGN activity), we
don't expect the gas temperature to be much higher than that of dust.

Ratios $r_{31}<0.4$ such as that found at the offset ($-7''$, $-42''$)
where $r_{31}$=0.19, indicates that the gas is either very cold ($\sim
10$ K), or of low density (less than 1000 \cm3). For 
$r_{31}$=0.2--0.4 typical LVG fits yield \Tk=10--20 K, $n=1000-2500$
\cm3, $\Lambda=6-12 \times 10^{-6}$ \XdVunit, and $X$ = $2-3 \times
10^{20}$ \Xunit, while $K_{\rm vir}$ is in the range of 3--4,
i.e. close to the virial value and thus consistent with what is
expected for typical ensembles of cold, self-gravitating molecular
clouds found in the Galaxy.  In general, without internal heating
sources, (\eg \, no star formation) typical gas temperatures in
molecular clouds are $\sim 10-15$ K, set by the equilibrium reached
between heating by ambient interstellar radiation ($G_{\circ}\sim 1-5$
in Habbing units) and comic rays and C$^+$ cooling, while they usually
have $ K_{\rm vir}\sim 1$ (to within a factor of a few).

 On the other hand, if we were to set \Tk = 30 K, the best fit would yield
 $n(\rm H_2) \sim 1000$ \cm3, and $\Lambda = 2-9 \times 10^{-6} $
 \XdVunit, which gives $X$ = $2.4-4.2 \times 10^{19}$ \Xunit. In this
 case, $K_{\rm vir}$ is in the range 9--30, too high for a quiescent
 galaxy similar to the Milky Way.  In \S 3.3.3, we will see that
 variations in the SCUBA $S_{450}/S_{850}$ flux density ratios
 indicate a variation in the dust temperature, with $T_{\rm dust}$
 much higher in the central region than in the outer disk.  Moreover
 the {\it global} spectral energy distribution (SED), with more data
 points available, is best fitted by a two-component dust model with
 $T_{\rm W}=27.9$ K and $T_{\rm C}=10.6$ K for the warm and the cold
 dust component respectively.  The warm dust component is then
 expected to be associated with the molecular gas in the nuclear
 region, while the cold dust with $T_{\rm C}=10 - 15$ K with the gas
 in the disk and spiral arms.

\subsection{NGC\,3310 and NGC\,157: The dust continuum emission}

\subsubsection{Dust temperature and mass: the methods}

Using the data obtained from the JCMT archive, Stevens et al. (2005)
published the SCUBA maps of 14 galaxies, including NGC\,3310 and
NGC\,157. They fitted their SEDs with a two-temperature dust
model. Their results show that NGC\,3310 has the largest fraction of
cold dust among the 14 sample galaxies.  However, there appeared to be
a mistake in the SED model of NGC\,3310 presented in Figure 17 of
Stevens et al. (2005), in which the submm fluxes were set to an order
of a magnitude larger than the real ones, and thus their reported dust
masses and temperatures are in error.

We re-analyzed the SCUBA maps, using our data obtained in 2000 (\S
2.2) and also included archival JCMT observations of these galaxies
(1999 August 08, 2000 January 19, 2000 April 13 and 2003 May 20;
weather grade 2, 3, 1 and 1 respectively).  The total fluxes derived
(Table 6) are in good agreement with those of Stevens et al.
(2005). We then correct for $^{12}$CO(3--2) line contamination of the
850 $\mu $m fluxes using the formula $S_{\rm CO}= 0.53 I_{\rm CO}$ mJy
beam$^{-1}$ (K km s$^{-1})^{-1}$ (Zhu et al.  2007).  This yields a
$\sim$ 15\% flux correction for NGC\,3310 (central 30$''$) and a
8--20\% flux correction for NGC\,157.  We furthermore corrected the
fluxes for contamination of, predominantly thermal, radio emission.  
Radio continuum observations at 5 different frequencies between 57 MHz
and 4.85 GHz were available from NED for NGC\,3310 and data points at
1.4 and 5 GHz for NGC\,157.  For NGC\,3310 we fitted a combination of
a synchrotron and thermal radio spectrum to the observed radio data
points by varying the synchrotron spectral index and the synchrotron
and thermal flux densities at 1 GHz.  The thermal radio emission has a
fixed spectral index of $-0.1$.  Fits were possible with a synchrotron
spectral index between $-0.7$ and $-0.8$ and yielded extrapolated
radio flux densities between 29 and 48 mJy at 850 \um and between 25
and 44 mJy at 450 $\mu$m.  We adopted the mean value as a realistic
estimate.  For NGC\,157 a detailed fit of the radio spectrum was not
possible.  We estimated the range of the radio continuum
contaminations by fitting a combined synchrotron and thermal radio
spectrum with the steepest ($-1.1$) and the flattest ($-0.6$)
synchrotron spectral index compatible with the radio data points at
1.4 and 5 GHz and found the range to lie between 5 and 28 mJy at 850
\um and between 3.5 and 26 mJy at 450 $\mu$m.  As a realistic estimate
for the radio continuum contamination we used a synchrotron spectral
index of $-0.9$ and derived radio flux densities of 18 and 16 mJy at
850 \um and 450 $\mu$m.  For both galaxies we found that the conclusions
of our study are not affected by the uncertainty in the radio
continuum contamination of our 850 \um and 450 \um flux densities.
The observed and corrected fluxes
(for $^{12}$CO(3--2) and radio contamination) are listed in
Table 6.

We then fitted the dust SEDs with two models: (i) a classic
two-temperature dust model (hereafter 2T model) similar to that used
in Stevens et al. (2005) and (ii) a dust grain model of Desert et al.
(1990, hereafter DBP90 model) which contains Large Grains (LG,
$\beta$=2), Very Small Grains (VSG, $\beta$=1) and Polycyclic Aromatic
Hydrocarbons (PAH).  The 2T model uses two grey-bodies to account for
the warm and cold dust components of temperatures $T_{\rm W}$ and
$T_{\rm C}$ respectively and, in our case, was applied to the data
with wavelengths $\ge 60~ \mu$m.  Both components of the 2T model have
the same dust emissivity $\beta=2$ in k$_\nu \propto \nu^{\beta}$
(Dunne \& Eales 2001).  The DBP90 model was previously applied to
NGC\,1569 (Lisenfeld et al. 2002) and here we use the same fitting
procedure.  We adopt the values for the size distribution of the
different grains for which DBP90 achieved the best-fit for the solar
neighbourhood, except for the relative abundance of the various grain
components which we leave as a free parameter.

The use of the DBP90 model was motivated by the inability of the
typical 2-component grey-body fit to reproduce the submm ``excess''
observed in NGC\,3310 (see \S 3.3.2) without yielding an unreasonably
cold dust temperature and a very large associated dust mass.  In
Appendix A we show how we derive the interstellar radiation field
(ISRF) for both galaxies. The ISRF in NGC\,3310 is much stronger than
in NGC\,157, even higher than for the starbursting dwarf galaxy
NGC\,1569 where indications for a boosted VSG population have been
found (Lisenfeld et al. 2002, Galliano et al. 2003).
  
The dust mass $M_{\rm dust}$ in the 2T model was calculated as the sum
of the warm ($M_{\rm w,dust}$) and the cold ($M_{\rm c,dust}$) dust
using

\begin{equation}
M_{\rm w/c, dust}(M_{\sun}) =\frac{S'_{\lambda_{\rm w/c}} D^2}{\kappa _{\lambda_{\rm w/c}} 
B_\nu(T_{\rm w/c,dust})},
\end{equation}

\noindent
where $\lambda_{\rm w/c}$ is the reference wavelength for each
component taken as 100 \um for the warm dust and 850 \um for the cold
dust and $T_{\rm w/c,dust}$ the temperature of the warm and cold
component, respectively.  $B_\nu(T_{\rm w/c,dust})$ is the Planck
function at 100 and 850 $\mu$m, respectively, $S'_{\lambda_{\rm w/c}}$
the flux density at 100 and 850 \um (after correcting the flux at 850
\um for $^{12}$CO(3--2) and radio continuum contamination) and $\kappa
_{\lambda_{\rm w/c}}$ the dust opacity per unit dust mass, assumed to
be $0.77 \,{\rm cm^2 g^{-1}}$ at 850 \um (c.f., James et al. 2002;
Dunne \& Eales 2001) and $25 \,{\rm cm^2 g^{-1}}$ at 100 \um
(Hildebrand 1983).  In DBP90, the dust temperature of the LGs was
derived by fitting equation (5) to the model. We then derived the mass of
LG component from the flux at 850 \um using the same equation.  The
masses of the VSGs and PAHs were derived from the input mass
extinction coefficients, obtained by scaling the values in Table 2 of
DBP90.  The results are presented in Table 7 and Table 8 for the 2T
and the DBP90 model respectively.

\subsubsection{The dust properties, a study of extremes}

The dust SED of NGC\,3310 is very different compared to a typical dust
SED of a spiral (see e.g. Dunne \& Eales 2001). The unusual aspect of
it is an apparent ``excess'' emission at 850 \um (see Figure 4a),
%(e.g. Bendo et al. 2006), 
which manifests itself as a relatively shallow submm slope ($\simeq$
$\lambda^{-2.5}$), derived from the $S_{450}/S_{850}$ ratio of 4.9
(after correcting for CO and thermal radio contribution).  This is
similar to that of the starburst dwarf galaxy NGC\,1569, which has a
$S_{450}/S_{850}$ ratio of 4.4 (Lisenfeld et al.  2002).  By
comparison, NGC\,157 has $S_{450}/S_{850}$=6.0 (after CO and thermal
radio correction), and all the galaxies in the SCUBA Local Universe
Galaxy Survey (SLUGS) have $S_{450}/S_{850}$$> 6$ (Dunne et al. 2000).
The value of $S_{450}/S_{850}= 4.9$ for NGC\,3310 corresponds to
$\beta < 2$, and constitutes strong evidence of large-scale
differences of dust properties in starburst environments.  This has
been suggested earlier using 1.3\,mm imaging (Dumke, Krause, \&
Wielebinski 2004), and along with the very high molecular gas
excitation implied by the high CO line ratios, points to different ISM
properties over large scales in starbursts. The latter, apart from
being interesting in its own right, can have serious implications for
interpreting molecular line and dust continuum data from the numerous
starbursts that have now been identified in the distant Universe.
 
The need for a different dust SED model in NGC\,3310 is exemplified by
the unlikely results yielded by the classic 2T model (Figure 4a) in
which a temperature of $5.7$ K is derived for the cold component which
would then contain the bulk of the mass with $M_{\rm dust} = 2.2
\times 10^8$ \ms (see Table 7), resulting in an unrealistically low
GD mass ratio of about 5 within the inner $60''$, the
region where most of the dust emission originates (see Table 8 for the gas
mass).  The highest cold dust temperature which is compatible with the
error bars of our data is $11.8$ K. In this case the dust mass would
be $M_{\rm dust} = 2.8 \times 10^7$ \ms\ and the GD ratio = 43, still a
low value.
The GD ratio for this likely metal-poor object ought to be higher not
lower than the Milky-Way value of about 100--150.  In principle, an
underestimate of the gas mass could be responsible for the low GD
ratio. This has been suggested for the Large Magellanic Clouds
(Bernard et al. 2008) where cold HI and/or a low density H$_2$ phase
without CO might cause a low GD ratio.  We cannot exclude this
possibility in NGC\,3310, but there are further, stronger arguments
against the presence of very cold dust.  A first argument is that the
extremely low dust temperature is also not consistent with the results
of our CO line excitation analysis. Based on the multi-transition CO
data, NGC\,3310 has a higher gas excitation, and a higher kinetic
temperature ($T_{\rm kin} > 30$ K) in the molecular gas content than
NGC\,157 ($T_{\rm kin} = 10 -20$ K). Thus, we would also expect a
higher average dust temperature, of the order 30 K or higher, in NGC \,3310
than in NGC\,157, which is an expected difference in the ISM
properties of an intense starbursts such as NGC3310.

A further argument against a very cold dust component is posed by the
question where the cold dust could hide.  Even though such low dust
temperatures can be achieved in very shielded environments, the dust
capable of producing this shielding would emit at much higher
temperatures.  Lisenfeld et al.  (2002) estimated that in the
case NGC\,1569 a considerable amount of (warm) dust is necessary to
shield the cold dust, so that an extreme dominance in mass of the cold
dust is impossible to happen.  Fischera \& Dopita (2008) modeled the SED
of dust emitted from self-gravitating, spherical, interstellar clouds
which allows to base such an estimate on a quantitative foundation.
The dust properties that they assumed are consistent with Galactic
dust.  They showed that massive clouds, close to the collapsing limit,
in a high pressure environment reach the highest opacities.  Although
in such clouds the temperatures of the grains in the inner part can be
below 10 K (see their Figure 8), the emission of this cold dust
component does not lead to a flat submillimeter spectrum (see their
Figures 10 and 11) since the overall emission of the cloud is still
dominated by the significant mass of warm grains from the outer
regions (needed to shield the inner ones so that their temperature can
drop that low).  Thus, very cold dust, even when present in the
centers of some clouds, is not expected to dominate the dust emission
for an ISRF similar or stronger than the Galactic one.

 Using the DBP90 model for NGC\,3310 on the other hand gives much more
 reasonable results (see Figure 5a), with a LG dust temperature of
 32\,K, but with an enhanced VSG abundance (factor 3.2) and a slightly
 lower PAH abundance (by a factor of 1.4) compared to the Galaxy (see
 Table 8). The enhancement of VSGs is similar to that found for the
 dwarf starburst galaxy NGC\,1569 (Lisenfeld et al.  2002) which is
 not unexpected given the fact that both galaxies are low-metallicity
 starbursts.  A submm excess and enhancement of VSG have also been 
 found for other starbursting dwarf galaxies (Galliano et
 al. 2005). The high abundance of VSGs may be the result of grain
 fragmentation due to numerous SN-induced shocks taking place in
 the ISM.  The total dust mass estimated with this model is $2.7
 \times10^6$ \ms, with the largest contribution coming from the 
 LG component ($2.1 \times10^6$ \ms).  The average dust temperature deduced
 for the LGs is 32\,K and is consistent with that of the
 H$_2$ gas.  The GD mass ratio within the central $60''$ (the region
 where most of the dust emission emanates from) is 461. This value is
 $\sim 3$ times higher than the Galactic value, but within the range
 expected for the low metallicity reported for NGC\,3310 (see \S
 3.4.1).
 
The good fit to the shallow submm SED of NGC 3310 is 
partly due to the fact that the stochastically heated VSGs possess
 a broad temperature distribution going down
to very low values. Most importantly, however, the 
low dependence of the extinction
coefficient on the wavelength ($\beta = 1$) of the VSG component
in the DBP90 model  is responsible for  the good fit.
The empirical nature of the DBP90 leaves some uncertainty about whether
this low value of $\beta$ is correct. In any case, a firm conclusion from the  
 good fit of the DBP90 model together with the impossibility
 to use  a two-temperature model (with $\beta=2)$  is   that  
 we need a low value of  $\beta$  in order to fit the shallow submm
spectrum of NGC 3310.

Low values of $\beta$ are expected for amorphous carbon (Koike et al
 1980), aggregates of silicates and graphite in a porous structure
 (Mathis \& Whiffen 1989) as well as for small amorphous grains (Seki
 \& Yamamoto 1989, Tielens \& Allamandola 1987).  In general,
 amorphous rather than crystalline structure and increase of the grain
 size (e.g.  Mannings \& Emerson 1994) are the main reasons for
 expecting $\beta <2$.  Furthermore, laboratory experiments have shown
 that $\beta$ can be temperature dependent and decreases at very low
 temperatures (Agladze et al. 1996) which could indicate that dust in
 the inner parts of dense clouds might show different properties than
 dust outside. Going into a similar direction, Meny et. al. (2007)
 proposed a dust model based on physical properties of amorphous
 solids which exhibits a broad emission spectrum whose detailed
 properties dependent on temperature and which might explain the submm
 excess.  Thus, an alternative explanation for the flat submm SED
 could be the change of dust properties with temperature.

The SED of NGC\,157 is similar to that of our Galaxy and both dust SED
models can fit the data well, but with a dust mass different by a
factor of 3--4.  From the 2T model we derived dust temperatures of
28\,K and 11\,K for the warm and the cold dust component
respectively (see Figure 4b and Table 7), in good agreement with the
results of Stevens et al. (2005).  The dust mass is $ 2.34\times 10^8
$ \ms, corresponding to a global GD mass ratio of 124,
similar to the Galactic value.  From the DBP90 model (Figure 5b) the
temperature derived for the large grains is 22.5\,K, with a total mass
of $ 6.2\times 10^7 $\ms ~(Table 8), and we achieved a good fit for
VSGs and PAH abundances similar to those in the Galaxy. This model
yields a GD mass ratio of 468, which is 3 times higher than the
Galactic value. This is mostly due to fitting the LGs with a single
temperature component of 22.5 K.  As shown in \S 3.3.3, the flux
density ratios of $S_{450}/S_{850}$ suggest that the dust temperature
indeed varies by a factor of 2 across the disk of NGC\,157, thus the
two-temperature model may better reflect the dust distribution in
normal spiral galaxies such as NGC\,157.

\subsubsection{The $S_{450}/S_{850}$ ratio: 
dust temperature variations, and VSGs versus LGs}

Having determined the proper dust model for each galaxy we can now
examine whether there are any discernible variations of the dust
temperature within these two systems.  Such variations have been
identified via submm imaging in the starburst/AGN NGC\,1068 by
Papadopoulos \& Seaquist (1999) using the $S_{450}/S_{850}$ ratio.
This ratio can be a good indicator for temperatures of $T_{\rm dust}
\leq 30$K, which characterize quiescent H$_2$ or \ion{H}{1}-rich
regions with little ongoing star formation.  For a single dominant
dust component this ratio can be expressed as

\begin{equation}
\frac{S_{450}}{S_{850}}=1.88^{\beta+3}(\frac{e^{16.8/T_{\rm dust}}-1}{e^{31.8/T_{\rm dust}}-1}),
\end{equation}

\noindent
with $\beta$ being the emissivity index (2 in our case).  For a
quiescent spiral such as NGC\,157, the $S_{450}/S_{850}$ variation
reflects a change in the dust temperature.  In Figure 6 we show the
azimuthally averaged $S_{450}/S_{850}$ ratio (after smoothing the
450$\mu $m map to the resolution of the 850$\mu $m map; Figure 6a) and
dust temperature profile as a function of radius (Figure 6b) for both
galaxies. We don't expect that equation (6) can be applied to
NGC\,3310, given that in this galaxy: a) the large grains have $T_{\rm
dust}\ga 30\,K$ (and thus their submm emission lies in the Rayleigh
Jeans domain), and b) the VSG component could be present throughout
the compact submm emitting region of this galaxy.  To demonstrate
this, Figure 6b shows that the ``dust temperature'' of NGC\,3310
derived with equation (6) is unreasonably low, much lower than that of
NGC\,157.  Therefore, the variation of the $S_{450}/S_{850}$ ratio for
NGC\,3310 in Figure 6a, may not reflect a simple variation in $T_{\rm
dust}$, but variations of the ``effective'' submm-deduced $\beta$,
possibly due to a change of the VSG/LG mass ratio, something that our
global dust SED fits of this system cannot discern.  We will address
this issue in a future paper, with detailed analysis of MIPS data from
Spitzer.

In NGC\,157, the dust temperature changes from 23--28 K in the central
region to less than 15 K in the outer disk (Figure 6b).  This is
consistent with the variation in the CO gas excitation temperature
estimated from the $r_{31}$ ratio, and again suggests that a
two-temperature dust model represents a good fit to the global SED in
this galaxy.

\subsection{CO, \ion{H}{1}, and dust distributions: an empirical test of the $X$ factor}

In this section we study the variation of the GD mass ratio in
NGC\,3310 and NGC\,157 in order to explore whether we can assume a
constant GD mass ratio for the \ion{H}{1}-dominated and
CO/H$_2$-dominated regions to get an independent, empirical test of
the $X $ factor (for a similar method applied in the Milky Way see
Reach et al.  1998).  The dust mass is estimated using equation (5)
and the \ion{H}{1} and H$_2$ mass can be estimated using the
following formulae:

%\begin{displaymath}
\begin{equation}
M_{\rm HI} (M_{\sun}) = 2.36 \times 10^5 D^2 S_{\rm HI}
\end{equation}
%\end{displaymath}
\begin{equation}
M_{\rm H_2}(M_{\sun}) = 1.073 \times 10^4 (X/X_{\rm G}) D^2 S_{\rm CO}, 
\end{equation}

%\begin{equation}
%M_{dust}(M_{\sun}) =\frac{S_{850} D^2}{\kappa _{850} B_\nu(T_{d})}
%\end{equation}
%\end{displaymath}
\noindent
where $D$ is the distance to the galaxy in Mpc, $ S_{\rm HI}$ and
$S_{\rm CO}$ is the \ion{H}{1} and $^{12}$CO(1--0) integrated flux in
Jy km $\rm s^{-1}$.  Equation (8) for the $M_{\rm H_2}$ estimate is
essentially equation (4) in Braine et al.  (2001) when using the
Galactic $X$ factor $X_{\rm G} $ = $2.0 \times 10^{20}$ $\rm cm^{-2}$
\rm (K km $\rm s^{-1})^{-1}$ or 4.3 \ms pc$^{-2}$ \rm (K km $\rm
s^{-1})^{-1}$ (including a 36\% mass correction for helium).  Our LVG
analysis shows that the factor $X/X_{\rm G}$ is approximately unity in
the disk of NGC\,157, but could be 0.65 in NGC\,3310 and 0.28 in the
nuclear region of NGC\,157.

For both galaxies, the $^{12}$CO(3--2) maps are more extended than
those of $^{12}$CO(1--0) and cover the majority of the CO
distribution.  They have higher S/N ratios and hence we used them to
represent the CO distribution, bearing in mind the variation of
$r_{31}$.  Also, we used the uncorrected SCUBA image to get the
highest S/N in order to trace the dust emission to the largest extent.
The CO contribution is on average about 15\% and thus not significant.

\subsubsection{ NGC\,3310 }

Figure 7a shows the $^{12}$CO(3--2) contours overlaid on the SCUBA 850
\um image while Figures 7b and 7c show the SCUBA 850 \um and
$^{12}$CO(3--2) contours respectively overlaid on the \ion{H}{1} map at
20\arcsec resolution, taken with the WSRT interferometer (Kregel \&
Sancisi 2001). The $^{12}$CO(3--2) map covers the majority of the CO
emission in the central $50'' \times 50''$ region, while the dust
emission can be detected out to a $60'' \times 90''$ region, thanks to
the high sensitivity of SCUBA. The IRAM $^{12}$CO(1--0) and
$^{12}$CO(2--1) maps published by Mulder et al. (1995) indicated the
existence of weak CO emission at $\pm 40''$ off the center, slightly
larger than the area we detected.  The total $^{12}$CO(1--0) flux from
the 45-m map is 101 Jy\,\kms.  Comparing to the total flux of 140
Jy\,\kms\ reported by Young et al. (1989), our map covers 72\% of the
CO, mostly in the central region.

The \ion{H}{1} emission is much more extended than the CO and SCUBA 850
\um emission and it dominates the gas content outside $60''$.  Unlike
NGC\,157 (see \S 3.4.2), there are no regions that are dominated by
CO.  In the central $20''$, the $M_{\rm HI}$/$M_{\rm H_2}$ ratio is
about 0.8 if $X=1.3 \times 10^{20}$ \Xunit is assumed.  However, the
\ion{H}{1} map at 20$''$ resolution is missing fluxes by about 30\%,
which was estimated by comparing to the 60\arcsec map (also from
Kregel \& Sancisi 2001), after convolving to the same resolution.
Thus we apply a factor of 1.3 to the high resolution \ion{H}{1} map,
and the resulting $M_{\rm HI}$/$M_{\rm H_2}$ is revised to 1.03,
\ie ~ neither \ion{H}{1} nor H$_2$ dominates the gas content in the
nuclear region.  The 850 \um flux is 67 mJy in the central $20''$, and
we estimate the dust mass by scaling this flux with the total flux to
obtain the fraction of $M_{\rm dust}$, derived from the DBP90 model,
in this region.  The final derived GD mass ratio is 312 (Table 9).  If
a Galactic $X$ factor is used, the central region would have 54\% more
molecular gas. The $M_{\rm HI}$/$M_{\rm H_2}$ and GD mass ratio
would become 0.67 and 395, respectively.

Similarly, in the central 60$''$ region ($r< 30''$), the GD mass ratio
would be 461 (the detailed numbers are listed in Table 9).  Our CO
data does not cover the entire 60$''$ region, but $M_{\rm H_2}$
contributes only about 27\% to the total gas mass, thus a 20-30\%
under-estimate in the CO fluxes would have a limited effect on the GD
mass ratio.

In the outer region with $r > 40''$ where little CO emission is
found, there is some weak 850\um emission associated with the
\ion{H}{1}.  To improve the S/N ratio, we averaged
the $S_{\rm 850}$ fluxes azimuthally in the outer disk  
and used the azimuthally averaged
$S_{\rm HI}/S_{\rm 850}$ ratio to estimate the GD mass ratio which 
is essentially the $M_{\rm HI}/M_{\rm dust}$ ratio as:

$$M_{\rm HI}/M_{\rm dust} =  593 (e^{16.8/T_{\rm dust}}-1)^{-1}.$$

\noindent
Using the value $T_{\rm dust} = 32$ K from the DBP90 model, we
estimate a GD mass ratio of 859.
If the VSG component is not enhanced in the
outer disk (possibly due to less star formation activities)
and $T_{\rm dust}$ drops to 20 K, the GD mass ratio would be 456.

In all cases, the resulting GD mass ratio in NGC\,3310 is much higher
than the Galactic value.  The high GD mass ratio could be related to
the low metallicity in the galaxy.  Pastoriza et al. (1993) found
metal abundances 2.5 to 5 times lower than Solar in circumnuclear
\ion{H}{2}  regions of NGC\,3310. It is thus possible that this
galaxy has less dust compared to a solar-metallicity galaxy.  The lack
of dust is consistent with the low global $M_{\rm H_2}$/$M_{\rm HI}$ ratio in NGC\,3310 as molecules are formed on the surfaces
of dust grains.

The GD mass ratio in NGC\,3310 shows larger variations than in NGC\,157 
(see \S 3.4.2). 
Also, due to the uncertainties in estimating the
fraction of the VSG component and hence the 
dust temperature, the derived local GD mass ratio has a large uncertainty,
by a factor of at least 2.  Combined with the fact that there are no
H$_2$ dominated regions in this galaxy, we conclude that it is not
practical to use the GD mass ratio to derive the $X$ factor in
NGC\,3310.

\subsubsection{ NGC\,157}

Figure 8a shows the $^{12}$CO(3--2) integrated intensity 
contours overlaid on a SCUBA 850 \um
image. The CO distribution is in general  consistent with that of the
dust emission. A major difference is visible in the nucleus
of the galaxy, where $^{12}$CO(3--2) emission is strongest but the dust
emission is relatively weak.  In fact, the strongest 850 \um  emission is
located in the southern spiral arm, close to the 
$^{12}$CO(1--0) peak (see Figure 2). 
This difference is probably due
to an excitation effect.  
The $^{12}$CO(3--2) is tracing hot molecular gas 
and its distribution is more
consistent with that of 450 \um emission (both have a strong emission
near the centre). This suggests that
different dust temperatures and $X$ factors (for different excitations)
should be used to derive the $M_{\rm H_2}/M_{\rm dust}$ ratios in different regions.

Figures 8b and 8c show the SCUBA 850 \um and $^{12}$CO(3--2) contours
respectively overlaid on the \ion{H}{1} map at a resolution of
$18''\times 12''$, obtained by Ryder et al. (1998) using the VLA.  The
\ion{H}{1} map was constructed using the uniform weighting schemes
(referred as UN image) which has a high resolution and shows the
structure of the inner region, but it misses some extended structure
hence is not good for deriving the \ion{H}{1} column density. We use it
to compare the \ion{H}{1} morphology with other ISM components in the
inner region.  Ryder et al. (1998) also published a low resolution
($41.5'' \times 27.2''$) VLA map processed with the nature weighting
schemes (referred as NA image), which is expected to recover all the
\ion{H}{1} fluxes.  We have converted the UN image to $42''\times 27''$
resolution to compare with the NA image and found that about 40-50\% of
fluxes could be missed in the UN image.

Based on these maps we divide the galaxy into 3 parts:
(i) the H$_2$ dominated region in the central $21''$,
(ii) the \ion{H}{1} dominated  outer disk with $\rm r>50''$
(iii) The regions between (i) and (ii).
In the following we discuss them separately.

(i) In the central $21''$ region, the $M_{\rm HI}$/$M_{\rm H_2}$
ratio is only 0.2 based on the \ion{H}{1} fluxes from the high
resolution UN image, after applying a correction factor of 2 for
missing fluxes.  The central region is clearly dominated by molecular
gas and thus the uncertainty in the \ion{H}{1} mass estimate has a
limited effect on the total GD mass ratio.  Line 4 of Table 9 lists
the mass ratios for the central 21\arcsec. For dust mass estimates we
used $T_{\rm dust}$=23 K which is the temperature we derived for the
central region of this galaxy (\S 3.3.3).  The GD mass ratio turns out
to be 54 if a low $X$ factor ($5.6 \times 10^{19}$ \Xunit) is used.

(ii) In the outer disk with $ r > 50''$, most of the gas are in the
atomic phase, and the GD ratio is effectively $M_{\rm HI}/M_{\rm
dust}$.  Similar to the case of NGC\,3310 outer disk, we used the
azimuthally averaged $S_{\rm HI}/S_{850}$ ratio to derive the $
M_{\rm HI}/M_{\rm dust} $ ratio as:

 $$M_{\rm HI}/M_{\rm dust} =  95 (e^{16.8/T_{\rm dust}}-1)^{-1} $$

\noindent
If $T_{\rm dust}$ = 15 K, as derived from equation (6) in \S 3.3.3,
the GD mass ratio would be 46. This is consistent with the GD mass
ratio in the central H$_2$ dominated region.

(iii) For the region with $10''< r < 50''$, we used the NA image to
derive the \ion{H}{1} fluxes and convolved the CO and 850 \um data to a
resolution of $42'' \times 27''$.  At such a low resolution, we select
three regions to derive the GD mass ratios: the center, the NE and the
SW disk that are 40$''$ offset from the nucleus. Table 9 lists the CO,
\ion{H}{1} and dust mass in these three regions and the GD mass
ratios. The $X$ factor was assumed to be $1.1 \times 10^{20}$ \Xunit
in the center, but $2.0 \times 10^{20}$ in the SW and NE spiral arms.
The dust temperature is 23 K in the center but 15 K at $40''$ away
from the nucleus, according to the temperature distribution in Figure
6b.

Our result indicates that the variation of the GD mass ratio is less than 40\%.
Thus we can assume a constant GD mass ratio of $62\pm 20$ to estimate 
the $X$ factor in the H$_2$ dominated region, which yields
$X = 6.6 \times 10^{19}$ \Xunit for the central $21''$, 
in good agreement, and independently, with the results from our LVG analysis
of the CO lines.

\section{Discussion}

\subsection{NGC\,3310 and NGC\,157: two extremes in ISM excitation}

Few molecular line ratio surveys of large samples of galaxies exist, and those
available are of the $r_{21}$ CO line ratio (Braine \& Combes
1992; Papadopoulos \& Seaquist 1998), while smaller samples have been observed
in  the more  excitation  sensitive  $r_{31}$ ratio  (Yao  et al.   2003;
Narayanan et  al.  2005).  With  the new array  receiver HARP and  the ongoing
JCMT  nearby  galaxies  legacy  survey (e.g. Wilson et al. 2009),  
more $r_{31}$  studies  of  nearby galaxies are expected.

In the Braine \& Combes (1992) survey the starburst NGC\,3310 and the
quiescent spiral galaxy NGC\,157 have respectively the highest and
lowest $r_{21}$ values ever observed in galaxies.  Although our
observations did not confirmed these extreme values, the $r_{31}$
ratio of 1.5 in NGC\,3310 remains among the highest of such ratios
measured in galaxies, while in NGC\,157, an extremely low ratio of
$r_{31}$= 0.2--0.4 is found in the outer disk.  These two galaxies
indeed represent two extremes in ISM excitation in external galaxies,
allowing us to study the range of the $X$
factor in extreme environments.

Our results showed an $X$ factor varying by a factor of $\sim 5$
between the two galaxies and within a galaxy where strong excitation
gradients are present.  In the disk region of NGC\,157 the $X$ factor
is close to the standard Galactic value found for Galactic GMCs at a
galactocentric radius of $ \sim 5-10$ kpc, while at its center it
could be lower by a factor of $\sim 5$, a trend also verified for the
Milky Way (Sodroski et al.  1995).  This good correspondence
demonstrates the utility of excitation analysis of multi-transition CO
data to yield good constraints on the $X$ factor in external galaxies,
even when a significant degree of degeneracy in the deduced average
physical conditions of the molecular gas remains.  The same method has
been used by Zhu et al.  to study the $X$ factor in interacting
galaxies, e.g.\, the Antennae (Zhu et al.  2003) and the Taffy
galaxies UGC 12914/15 (Zhu et al.  2007).  These galaxies represent
other types of extreme environments for molecular clouds, with strong
tidal interaction, direct cloud-cloud collision, and diffuse clouds
outside a galactic disk environment (in the case of Taffy
galaxies). The $X$ factor was found to be in the range of 0.1 -- 0.2
$X_{\rm G}$ in the these galaxies.  Yao et al.  (2003) also found on
average $X=0.1 X_{\rm G}$ based on their analysis of the $r_{31}$
ratios in the central regions of SLUGS galaxies.

In most  cases studied so far, a lower  than standard $X$ factor  is associated
with  non-virialized clouds  rather than  non-standard molecular  gas physical
conditions. According to equations (3) and (4)
\begin{equation}
X = \frac{n}{T_{\rm rad}} \frac{1}{dV/dr} = \frac{\sqrt{n}}{T_{\rm rad}} 
\frac{1.54}{\sqrt{\alpha}}  K_{\rm vir}^{-1} ,
\end{equation}
with $K_{\rm vir} = 1$ for ensembles of virialized clouds (the
ensemble itself in general does not need to be virialized, and its
motions ordinarily are dictated by the general galactic
potential). Most molecular clouds in a quiescent galaxy have $K_{\rm
vir}\sim 1$ and $X\sim X_{\rm G}$.  Our LVG excitation analysis find a
$\sqrt{n}/T_{\rm rad}$ varying at most within factors of $\la 3$,
while major variations come from $K_{\rm vir}>1$.  In the Milky Way,
very large local velocity gradients (greater than 10 km s$^{-1}$
pc$^{-1}$) with $ K_{\rm vir}\gg 1$ are found near pre-star-forming
clouds (Falgarone et al.  1998) as well as in diffuse clouds towards
Orion (Knapp \& Bowers  1988).  Falgarone et al. (1998) found high velocity
gradients associated with macroturbulent gas ``cells'' of size $< $
200 AU and densities of \nh2 $\sim 10^3$ \cm3, which is consistent
with the parameters found in our LVG analysis.

In the  nuclear regions  of ULIRGs, $K_{\rm vir}$ can be  significantly higher
than  unity,  due to  a  large  scale ``mixing''  of  molecular  gas  with
significant amounts  of stellar  mass, resulting in  $X \sim  0.2 X_{\rm G}$
(Solomon et  al. 1997; Downes \&  Solomon 1998). Actually our  method of using
the LVG radiative  transfer models to obtain $K_{\rm vir}$  values can in 
such a case (with gas and star  mixtures) yield  the factor
$f_{\rm g}=M_{\rm H_2}/M_{\rm total}=(\Delta V/\Delta V_{\rm vir})^2 = K_{\rm vir}^{-2}$, 
\noindent
independent of methods that  model the detailed spatial  and velocity
distribution of molecular gas in such systems, which demand high resolution
interferometric imaging (e.g. Downes \& Solomon 1998).

Summarizing our findings for the variations of the $X$ factors in NGC\,3310 and
NGC\, 157, as well as in the aforementioned ULIRGs and 
interacting galaxies (Zhu et al. 2003, 2007), 
we find that the $X$ factor is generally in the range of 
0.1--0.5 $X_{\rm G}$ in extreme environments,  which is
mostly due to  the variations of the parameter $K_{\rm vir}$ ranging
from 1 to $\sim$ 15.

\subsection{Molecular gas estimates in extreme environments: cautionary tales}

The  discovery of a  significant population  of dust-enshrouded  and CO-bright
starbursts at  high redshifts (e.g. Smail, Ivison, \& Blain  1997; Hughes et
al.  1998;  Greve et al.   2005), as well  as the discovery of  extended outer
regions dominated by cold dust (and concomitant gas) in several spirals
(Nelson,  Zaritsky, \&  Cutri 1998;  Xilouris  et al.   1999; Papadopoulos  \&
Seaquist 1999;  Thomas et al.   2001, 2002) give  to our study of  extreme ISM
extragalactic environments a wider importance.

For high-z galaxies for which multi-transition CO data are rarely
available we can only make an educated guess of the $X$ factor based
on our knowledge of local starbursts, frequently used as their low-z
counterparts based on the high star formation rates per molecular gas
mass found for them. A gas dynamics driven by fast-evolving galactic
potentials associated with mergers and/or frequent SNR shocks could be
the decisive similarity among them that makes claims about similarly
low $X$ factors in both populations credible.  Indeed if these low $X$
values are driven mostly by large velocity gradients rather than gas
excitation differences, highly turbulent environments would be
expected to be present in them.  However, it has been shown that
interstellar turbulence decays quite rapidly, on timescales of the
order of the free-fall time of the system (e.g.  Klessen et al. 2000; 
Mac Low et al. 1998).  To maintain the observed long lifetimes, turbulence in
molecular clouds must be constantly driven, either externally (e.g.
via galaxy interactions) or internally via star formation.  Therefore,
generally a gas-rich galaxy with a high star formation efficiency,
indicated by a high $L_{\rm FIR}/L_{\rm CO}$, may have systematically
lower $X$ factors.  Indeed, the $L_{\rm FIR}/L_{\rm CO}$ ratio is much
higher in NGC\,3310 and ULIRGs than in quiescent spirals. Tacconi et
al. (2008) shows an anti-correlation between the $X$ factor and the
gas surface density (see their Figure 10), implying that galaxies with
more dense gas and therefore higher star formation efficiencies would
have lower $X$ factors.  We will explore this effect more
quantitatively in a future paper.

The other method of deriving H$_2$ gas mass from the dust mass, a
benchmarking of the GD mass ratio from the \ion{H}{1}-rich part of the
galaxy, while independent of the $X$ factor and thus valuable, cannot
be currently applied to high redshifts galaxies. Its use is restricted
in the local Universe of $z\la 0.3$ where \ion{H}{1} can be imaged by
current interferometers.  Moreover, the presented evidence for
different dust properties in the intense starburst galaxy NGC\,3310,
while extremely interesting by themselves (especially if found
generally true for similar galaxies) complicates further the task of
estimating the H$_2$ gas mass following this method, even if
\ion{H}{1}, IR/submm dust continuum and CO images are available.

\section{Summary and Conclusions}

We present fully-sampled $^{12}$CO(3--2) and $^{12}$CO(1--0)
observations and 850$\mu$m and 450$\mu$m dust continuum imaging, all
at a matched beam resolution of $\rm HPBW\sim 15''$, for the starburst
NGC\,3310 and the quiescent spiral NGC\,157 that represent two
extremes of molecular gas and dust excitation conditions in
IR-luminous galaxies.  Including available sensitive \ion{H}{1} maps
convolved to a similar resolution, we have obtained a detailed view of all
phases of the neutral ISM in these galaxies, and an examination of the
standard method of using $^{12}$CO(1--0) brightness in deducing H$_2$
gas mass.  Our conclusions are summarized as the following:

(1) The  extremely high  $r_{21}$  in NGC\,3310  and low  $r_{21}$ in
NGC\,157  reported  by  Braine  et  al.   (1993)  are  not  confirmed  by  our
observations. The $r_{31}$ and  $r_{21}$ ratios in NGC\,3310 are still
at the high end;  while in NGC\,157 the $r_{31}$ ratio  is found to be low
only in the outer disk.

(2) NGC\,157 shows large variations in the $r_{31}$ ratio, which
is interpreted as variations in excitation condition and hence the $X$
factor. The $X$ factor is close to the Galactic value $X_{\rm G}$ in the
quiescent regions of  the outer disk where the temperature is cold (no
heating) and the velocity gradient is low.  In the nuclear region, on
the other hand, the temperature and gas velocity gradient are
high, resulting in an $X$ factor being $\sim 4$ times lower than
$X_{\rm G}$, assuming $Z_{\rm CO}$ is $5\times 10^{-5}$.

(3) The gas-to-dust mass ratio found for NGC\,157 is consistent with that
of the Galaxy, and using its benchmarked value in its \ion{H}{1}-rich
part (assuming it stays the same over its entire disk) independently
yields an $X$ factor of $\sim 4$ times lower than its Galactic value
in its H$_2$-rich nuclear region.

(4) The $X$ factor in NGC\,3310 can be $\sim 8$ times lower than its
Galactic value for typical CO abundances $Z_{\rm CO}=10^{-4}$.  If one
assumes that the lower metallicity values deduced for a few \ion{H}{2}
regions in this galaxy are representative of its bulk ISM
($Z_{\rm CO}=2\times 10^{-5}$), then the $X$ factor is only $\sim 2$ times
smaller than the Galactic values with the excitation effects mostly
offset by the lower metallicities.

(5) We deduce an unexpected and substantial contribution from a
potentially very different grain population in the starburst
NGC\,3310, manifesting itself as an observed submm ``excess'' which a
classic 2T model fit to its dust emission SED would otherwise
interpret as extremely cold dust with $T_{\rm C}\sim 5 - 11$ K which
corresponds to gas-to-dust mass ratios of $\sim 5-43 $, very unlikely for
a starbursting and most likely metal-poor system.  We find that a low
value for the frequency dependence of the extinction coefficient,
$\beta \sim 1$, is required to explain the observations.  We find that
a small grain population, a possible result of the extreme radiative
and SNR-dominated ISM environment of this extreme system, can account
for this submm ``excess'' without postulating very cold dust
temperatures in such an unlikely environment. We also discuss other
possible explanations.

\acknowledgements

The authors wish to thank the staff of the JCMT for their generous
assistance during an extended period of observations.  The support
personnel at the 45-m telescope in Nobeyama, Japan is also gratefully
acknowledged for their assistance.  We also thank Stuart Ryder and
Michiel Kregel for providing \ion{H}{1} images.  UL acknowledges
financial support from the research project AYA2007-67625-C02-02 from
the Spanish Ministerio de Ciencia y Educaci\'on and from the Junta de
Anaduc\'\i a. In addition, we thank the anonymous referee for stimulating
suggestions and comments.

\appendix

\section{The Interstellar Radiation Field of NGC\,3310 and NGC\,157}

In order to calculate the dust emission, in the concept of the DBP90
model, we need to quantify the Interstellar Radiation Field (ISRF) in
which the dust is immersed.  We derive this radiation field from the
integrated extinction corrected fluxes in the UV and optical
wavelength (listed in Table 10), assuming a spherical symmetry of the
emitting region with a radius of 3.2 kpc for NGC\,3310 and 12.7 kpc
for NGC\,157.  Both radii are estimated from the UV (GALEX archival
data reduced by us), optical and SCUBA maps.  The small size in
NGC\,3310 reflects the fact that both the UV and optical radiation and
the dust emission are very concentrated to the center. Figure 9 shows
the ISRF field for both galaxies (see the figure caption for
explanation of the different symbols). We can see that the ISRF in
NGC\,3310 is even stronger than that of NGC\,1569, which could enhance
the population of very small grains.

\clearpage

%%%%%%%%%%%%%%%%%%%%%%%%%%%%% FIGURES %%%%%%%%%%%%%%%%%%%%%%%%%%%%%%
\noindent
{\bf FIGURES:}

%Figure 1
\begin{figure}
\includegraphics[angle=0,scale=.8]{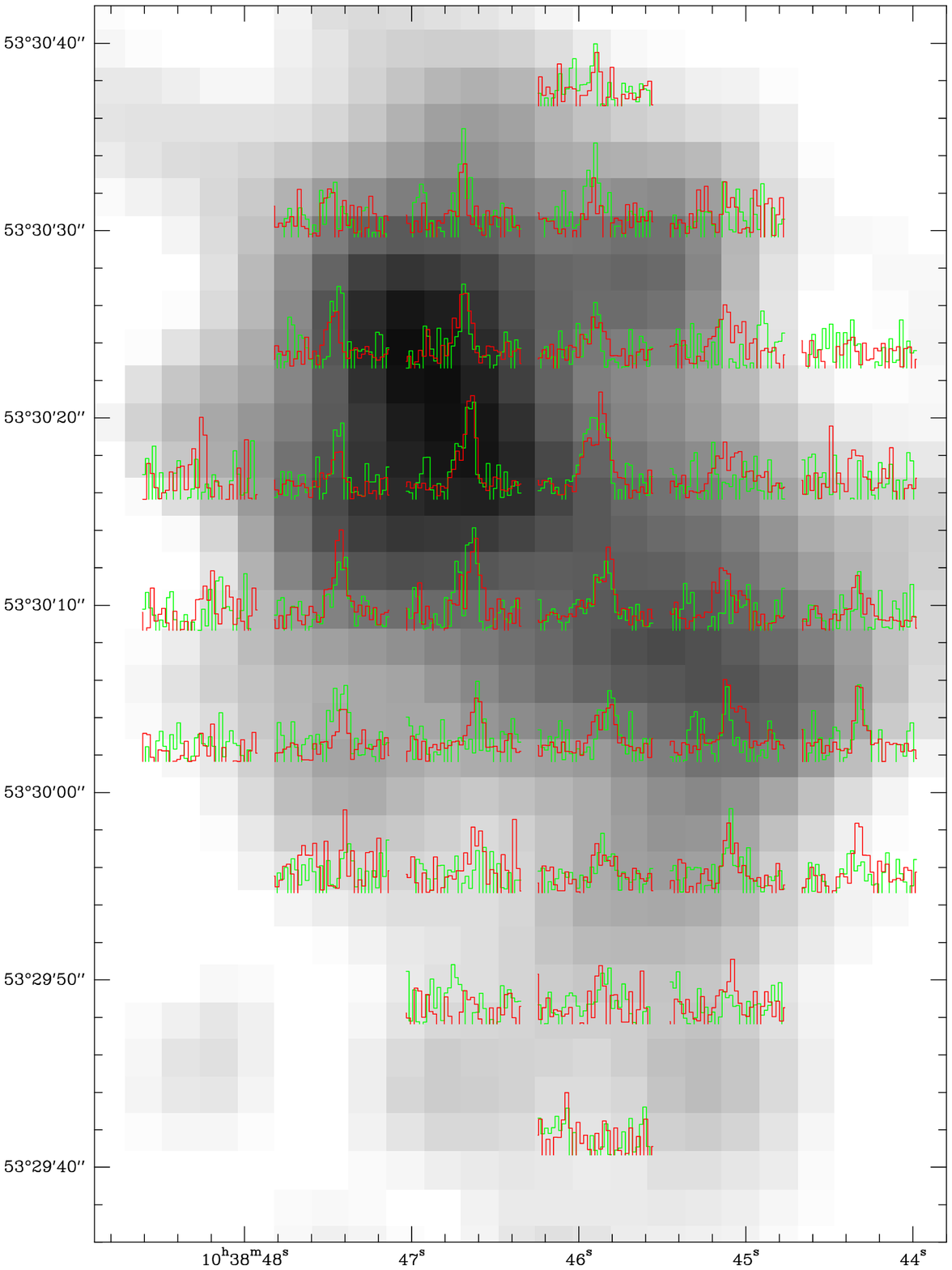}
\figcaption{ 
The $^{12}$CO(3--2) profiles (red) overlaid on the $^{12}$CO(1--0)
profiles (green) of NGC\,3310.
The background grey-scale map is the SCUBA image at $850 \mu$m.
The spectra have been smoothed to a spectral resolution of 21.6 \kms.
The velocity range is 660 to 1320 \kms. In each plot
the Y-axis is $T_{\rm mb}$ which ranges from -0.02 to 0.15 K.
\label{fig-1}}
\end{figure}

%Figure 2. 
\begin{figure}
\includegraphics[angle=0,scale=1.0]{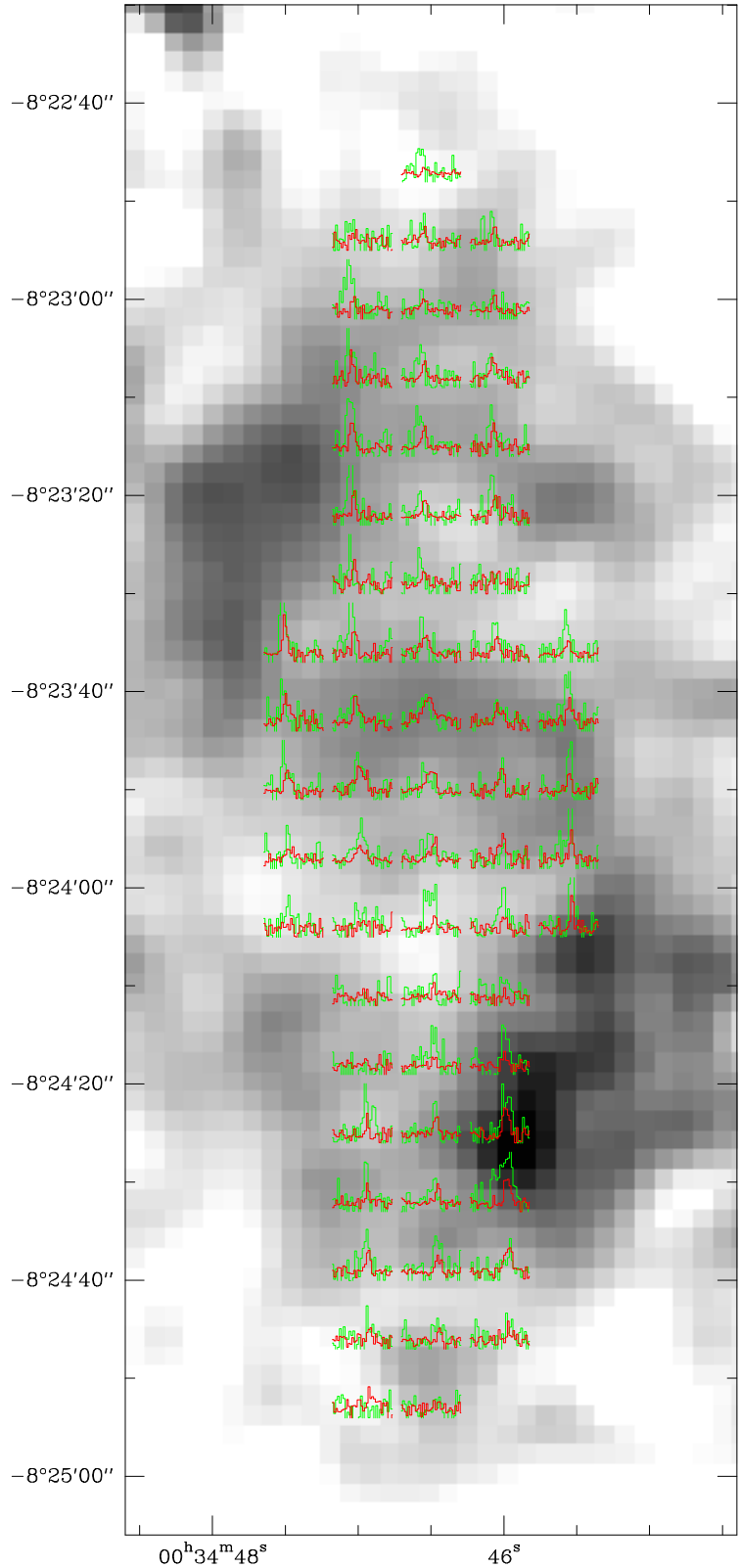}
\figcaption{ 
The $^{12}$CO(3--2) profiles (red) overlaid on the $^{12}$CO(1--0)
profiles (green) of NGC\,157. The background grey-scale map
is the SCUBA image at $850 \mu$m.
%(n157_co32_scuba.eps)
The velocity range is 1360 to 1996 \kms for $^{12}$CO(1--0)  
but 1300 to 2020 \kms for $^{12}$CO(3--2).
In each plot the Y-axis is $T_{\rm mb}$ which ranges 
from -0.05 to 0.18 K. 
\label{fig-2}}
\end{figure}

%Figure 3. 
\begin{figure}
\includegraphics[angle=0,scale=.5]{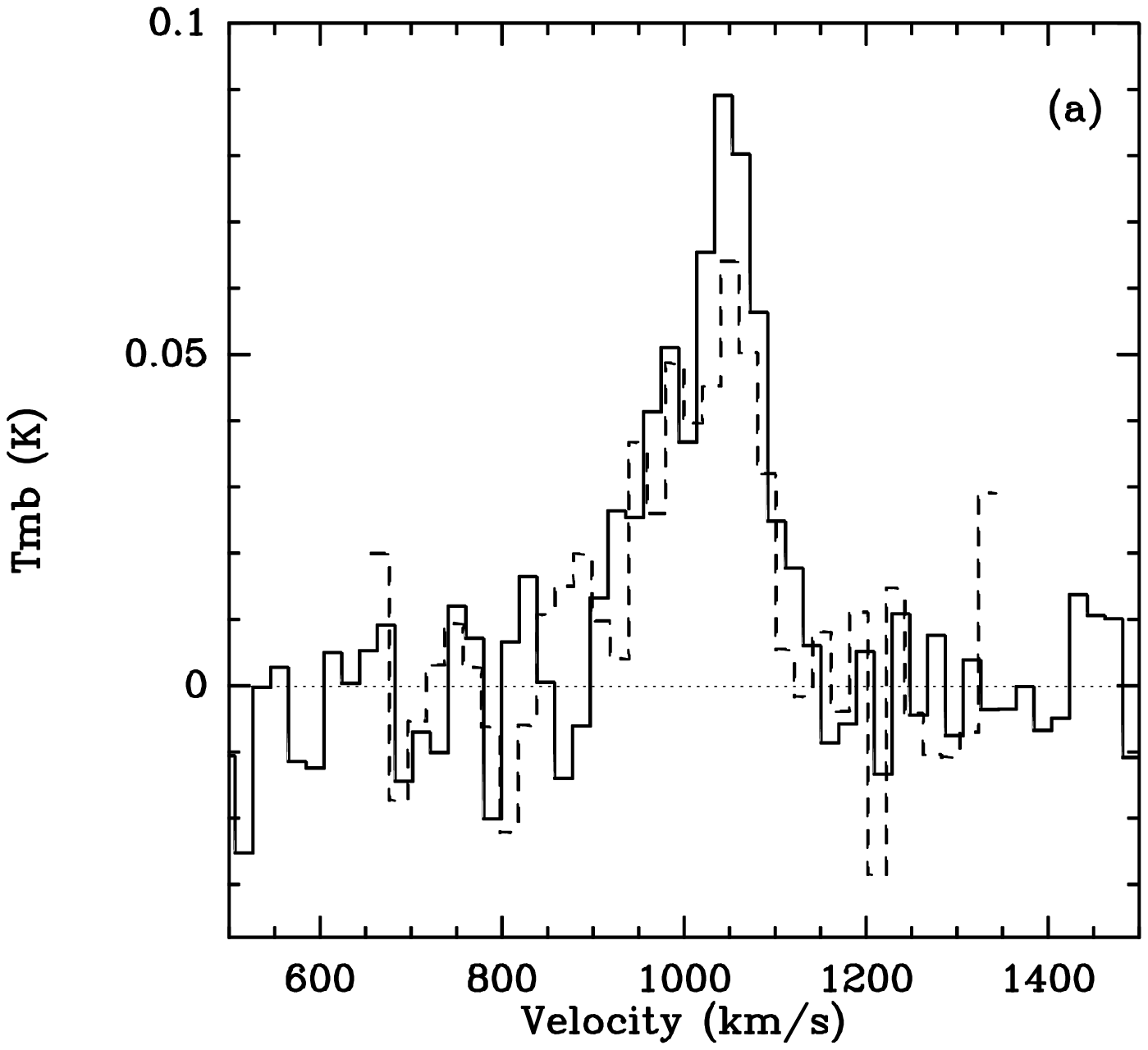}
\includegraphics[angle=0,scale=.5]{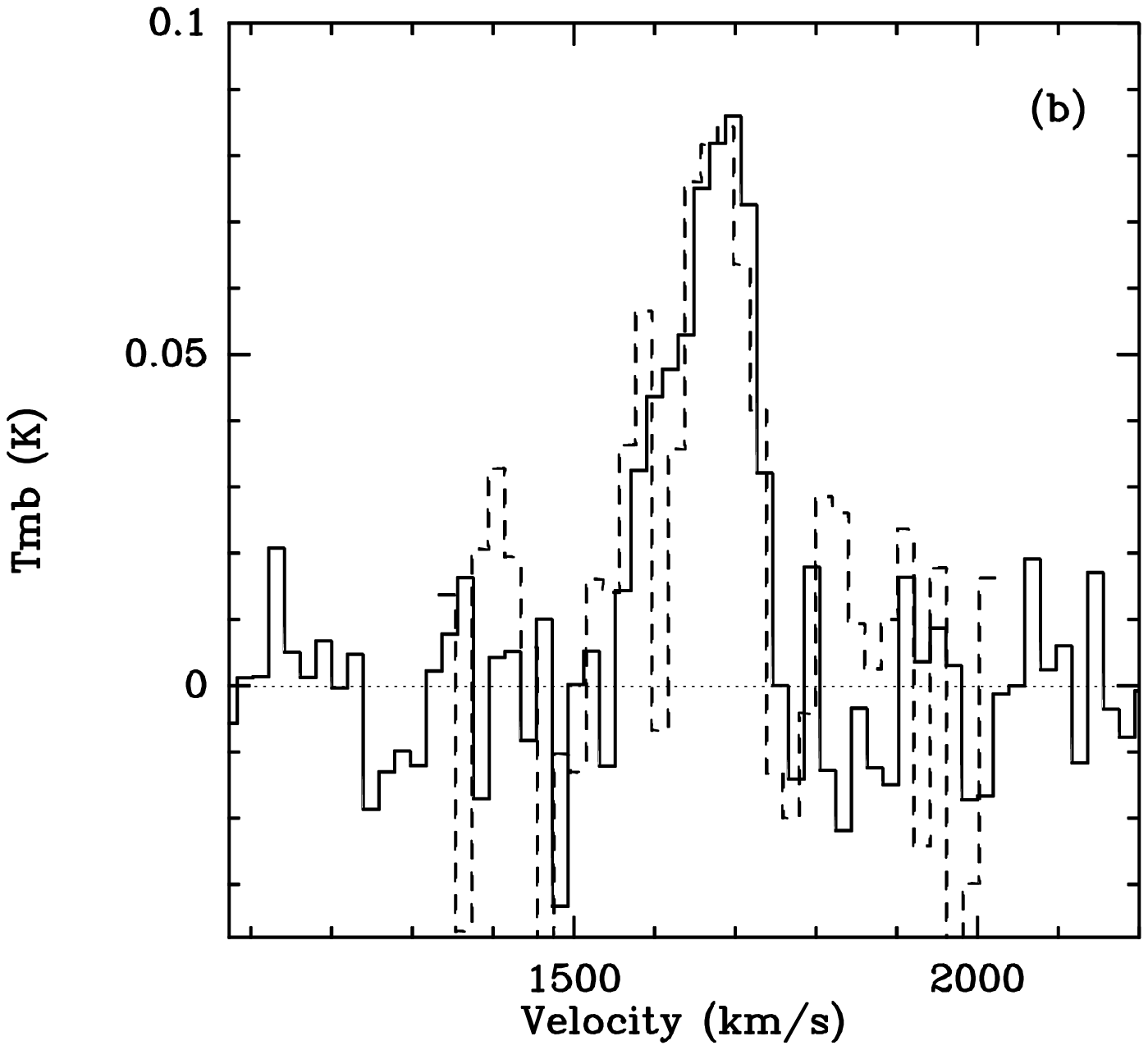}
\figcaption{
The $^{12}$CO(2--1) spectrum observed with the JCMT 
(solid line) overlaid on the $^{12}$CO(1--0) spectrum (dashed line) 
from the NRO 45-m for NGC\,3310 (a) and NGC\,157 (b). 
\label{fig-3}}
\end{figure}

%Figure 4
\begin{figure}
\includegraphics[angle=0,scale=.5]{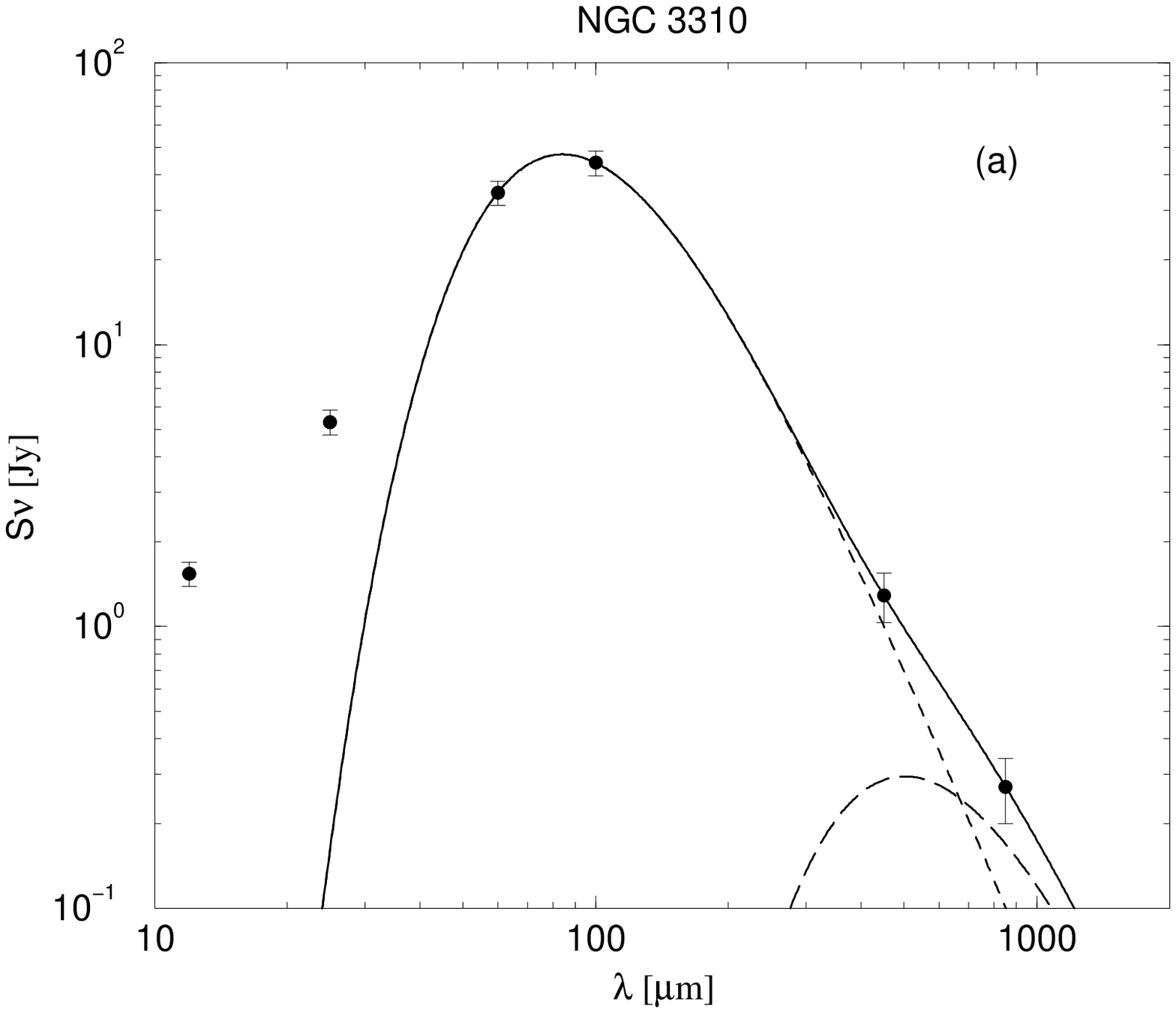}
\includegraphics[angle=0,scale=.5]{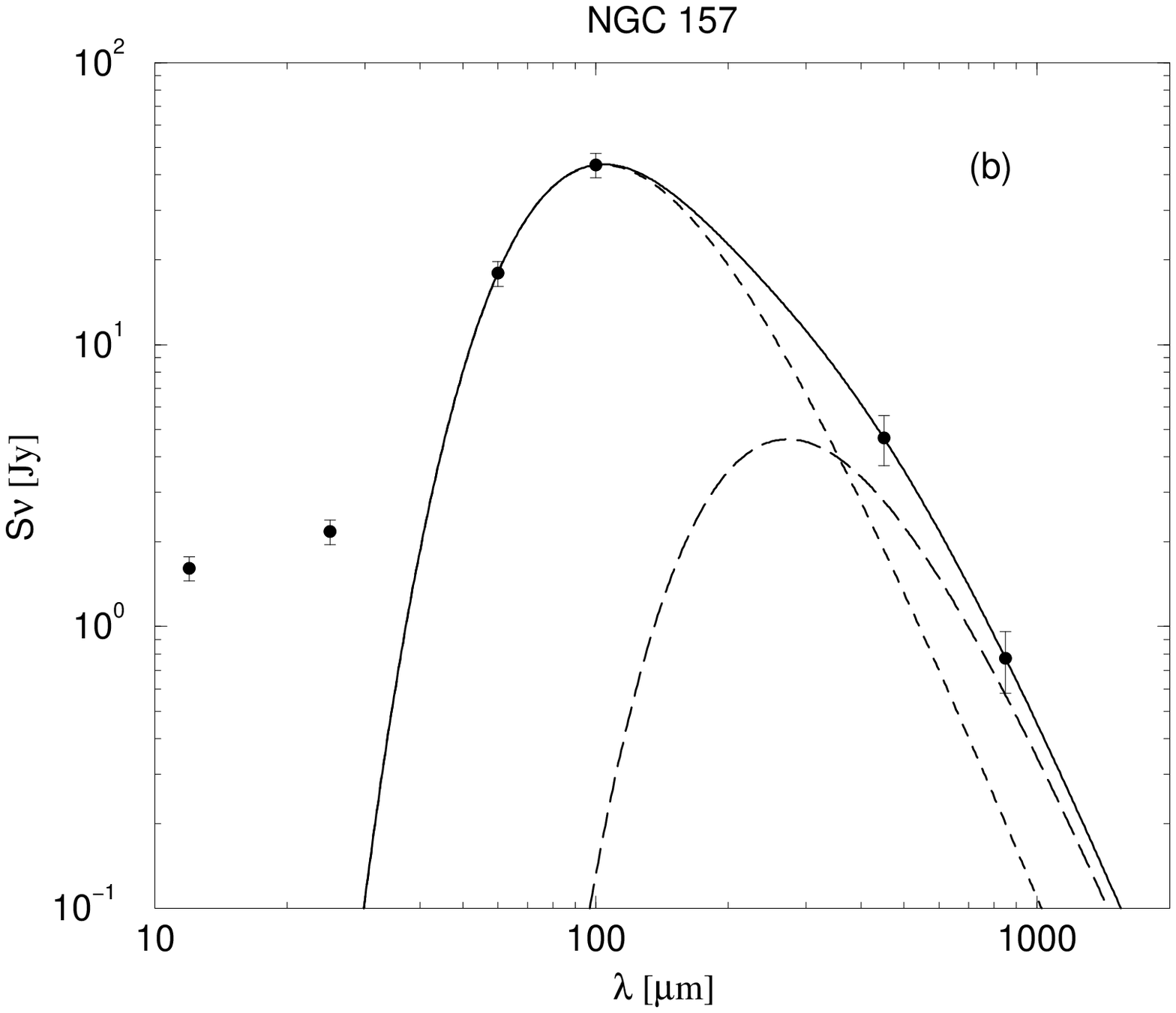}
\figcaption{ 
Two-temperature dust model fitting to the SEDs of the two galaxies
(a) NGC\,3310 and (b) NGC\,157. Two grey-bodies were 
fit to the data (for wavelengths $\ge 60$ \um) to account
for the warm and the cold dust components. 
\label{fig-4}}
\end{figure}

%Figure 5
\begin{figure}
\includegraphics[angle=0,scale=.5]{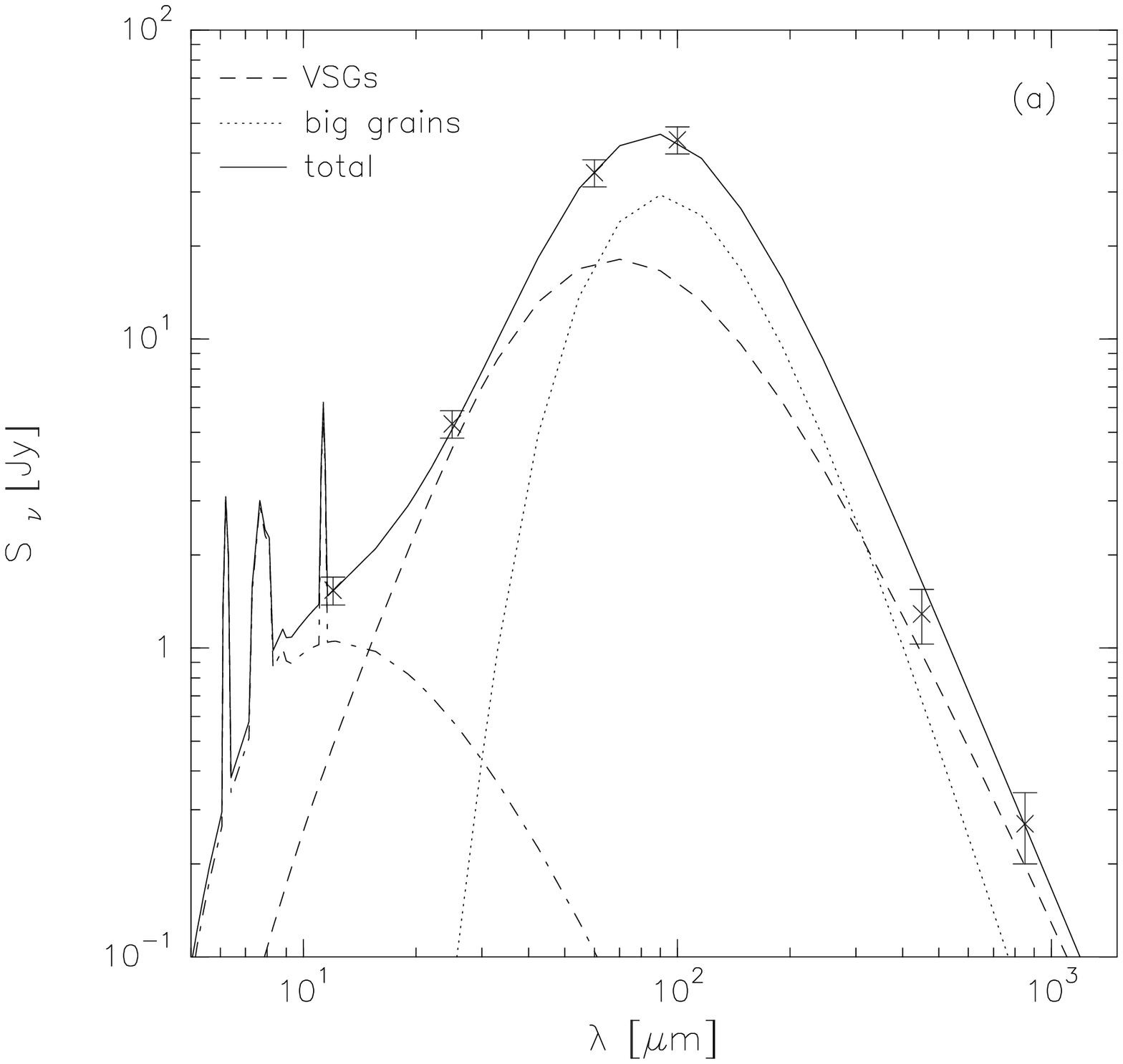}
\includegraphics[angle=0,scale=.5]{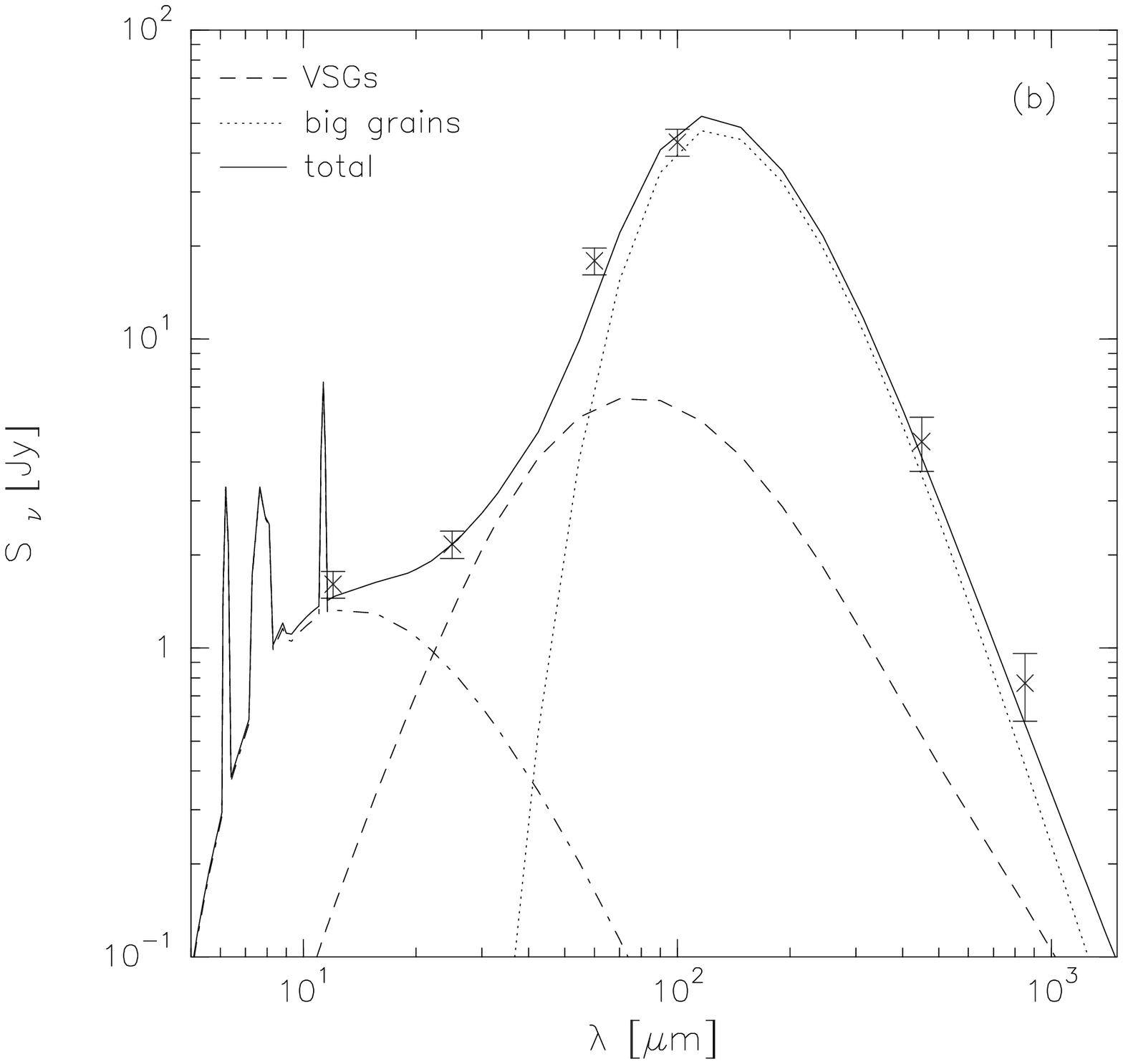}
\figcaption{
The SED fitting using the  Desert et al. (1990) dust model  for  
NGC\,3310 (a) and NGC\,157 (b).
\label{fig-5}}
\end{figure}

%Figure 6
\begin{figure}
\includegraphics[angle=0,scale=.5]{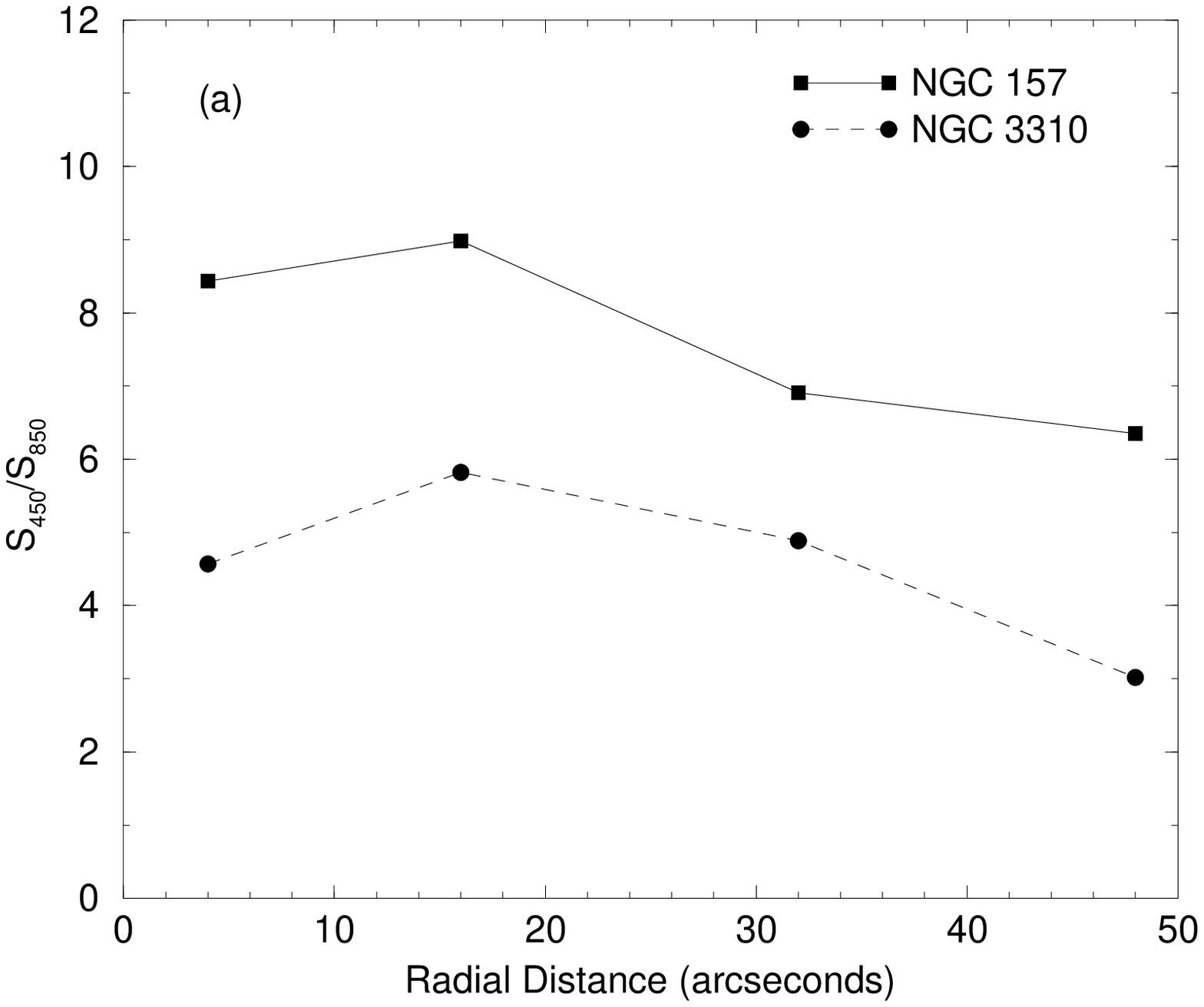}
\includegraphics[angle=0,scale=.5]{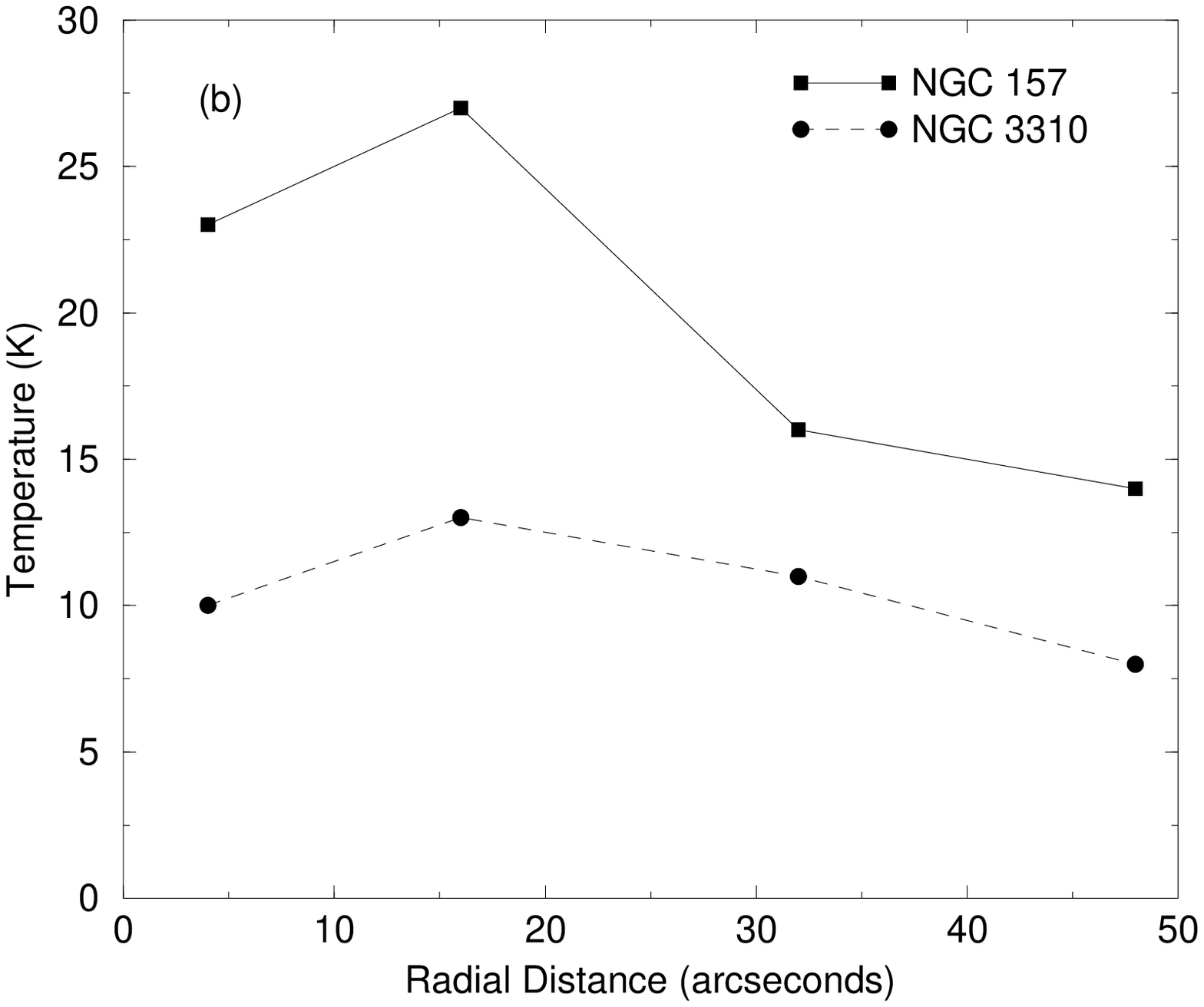}
\figcaption{
Azimuthally averaged profiles of 
(a) the $S_{450}/S_{850}$ flux ratio, and 
(b) the dust temperature derived from equation (6)
for NGC\,3310 and NGC\,157. A radial step size of $16''$,
comparable with the $850 \mu m$ beam size, was used.
\label{fig-6}}
\end{figure}
\newpage

%Figure 7
\begin{figure}
\includegraphics[angle=0,scale=.4]{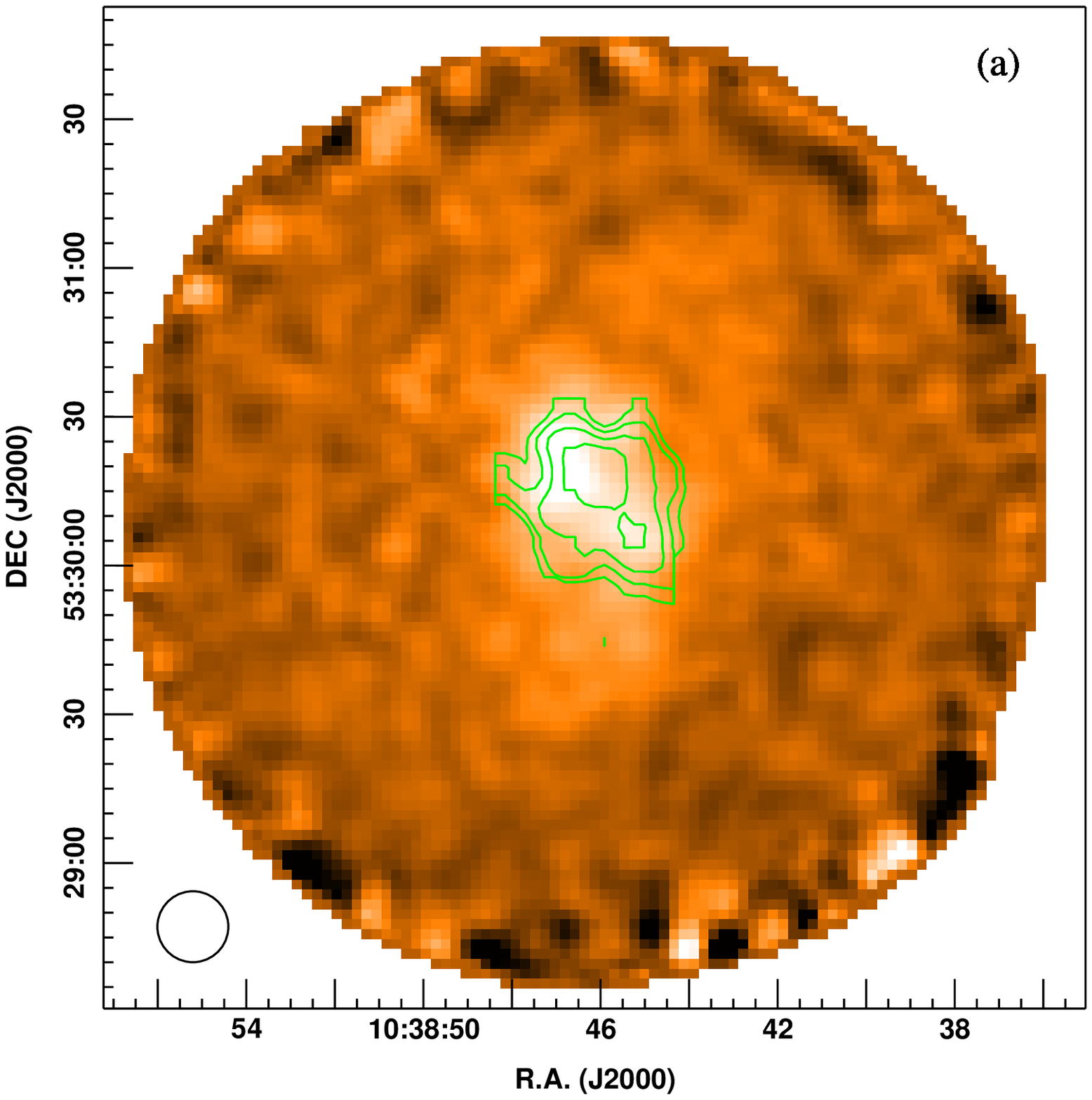}
\includegraphics[angle=0,scale=.45]{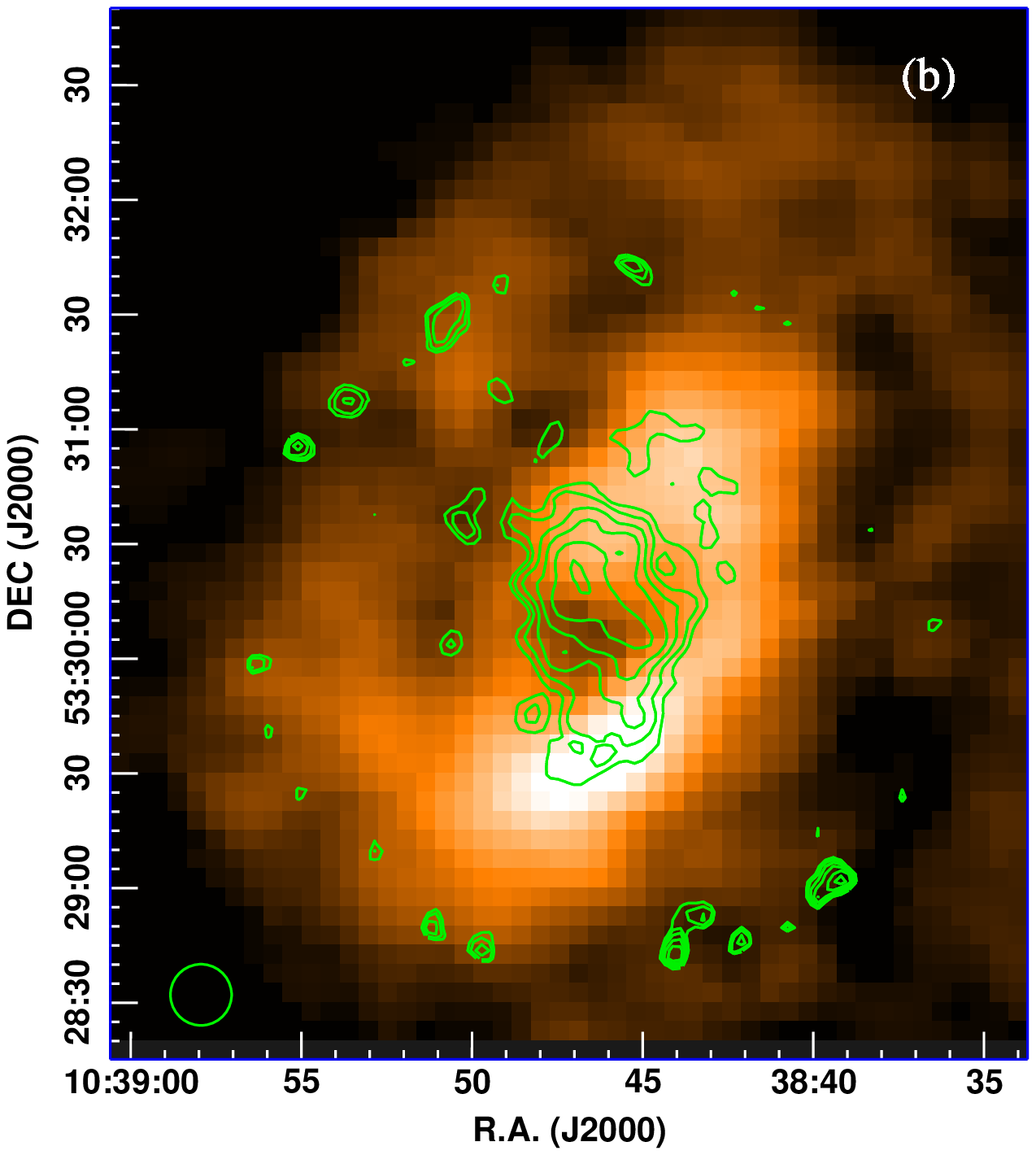}
\includegraphics[angle=0,scale=.425]{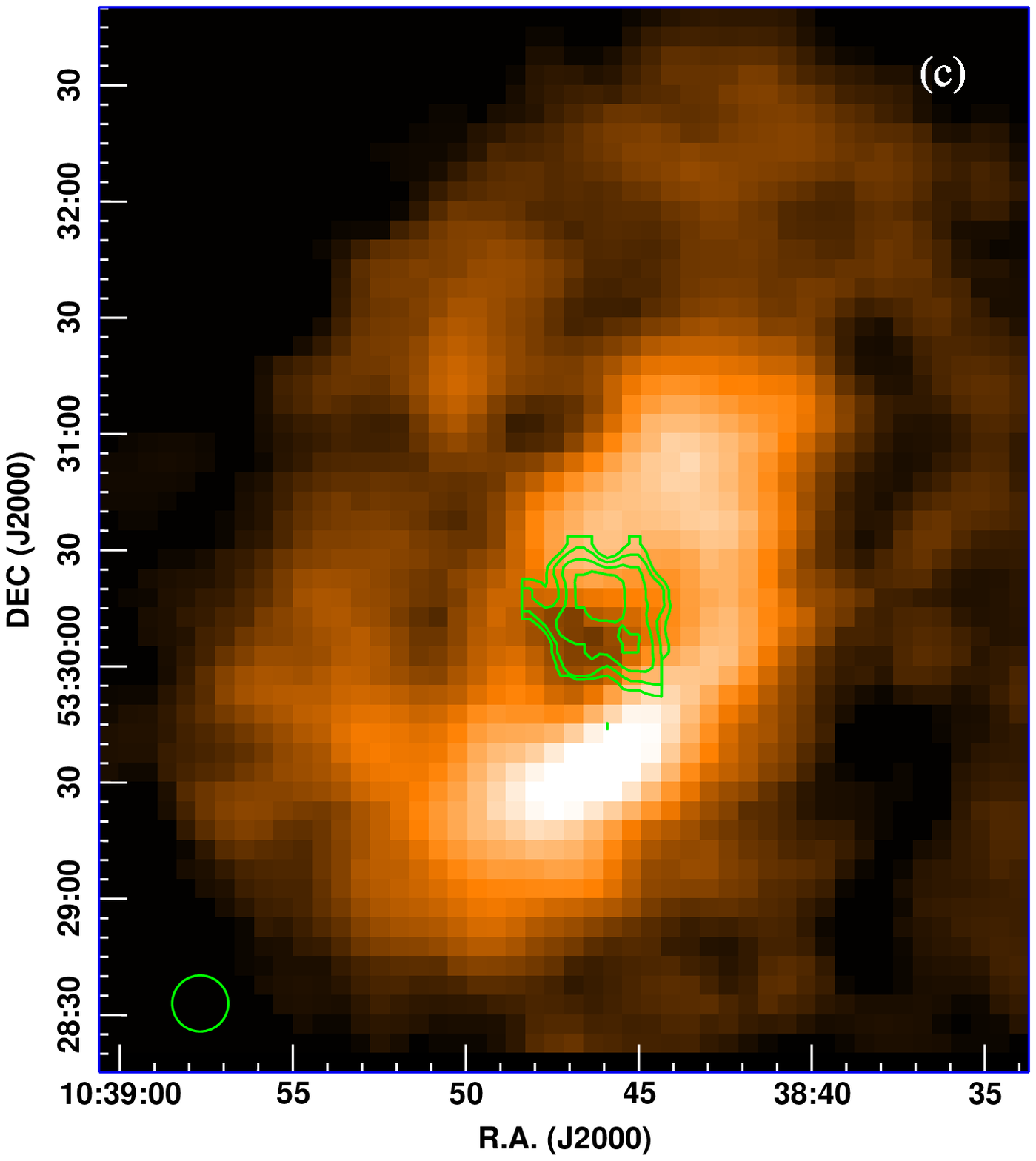}

\figcaption{
Morphological comparison between the different ISM tracers (\ion{H}{1},
$850 \mu m$ and $^{12}$CO(3--2)) for NGC\,3310. 
In Figure 7a, the $^{12}$CO(3--2) contours are overlaid on top
of the SCUBA 850 \um map. The contour levels are 4, 5.6, 8, 11.2 and 16 \kkms.
In Figure 7b, the 850 \um contours are overlaid on top of the 
the \ion{H}{1} map. The contour levels are 10, 14, 20 28, 40 and 56 mJy/beam.
Figure 7c presents the $^{12}$CO(3--2) contours on top of the
\ion{H}{1} map. The contour levels are as in Figure 7a.
In all maps the beam size ($15''$) is indicated on the bottom-left corner.
\label{fig-7}}
\end{figure}
\newpage

%Figure 8
\begin{figure}
\includegraphics[angle=0,scale=.4]{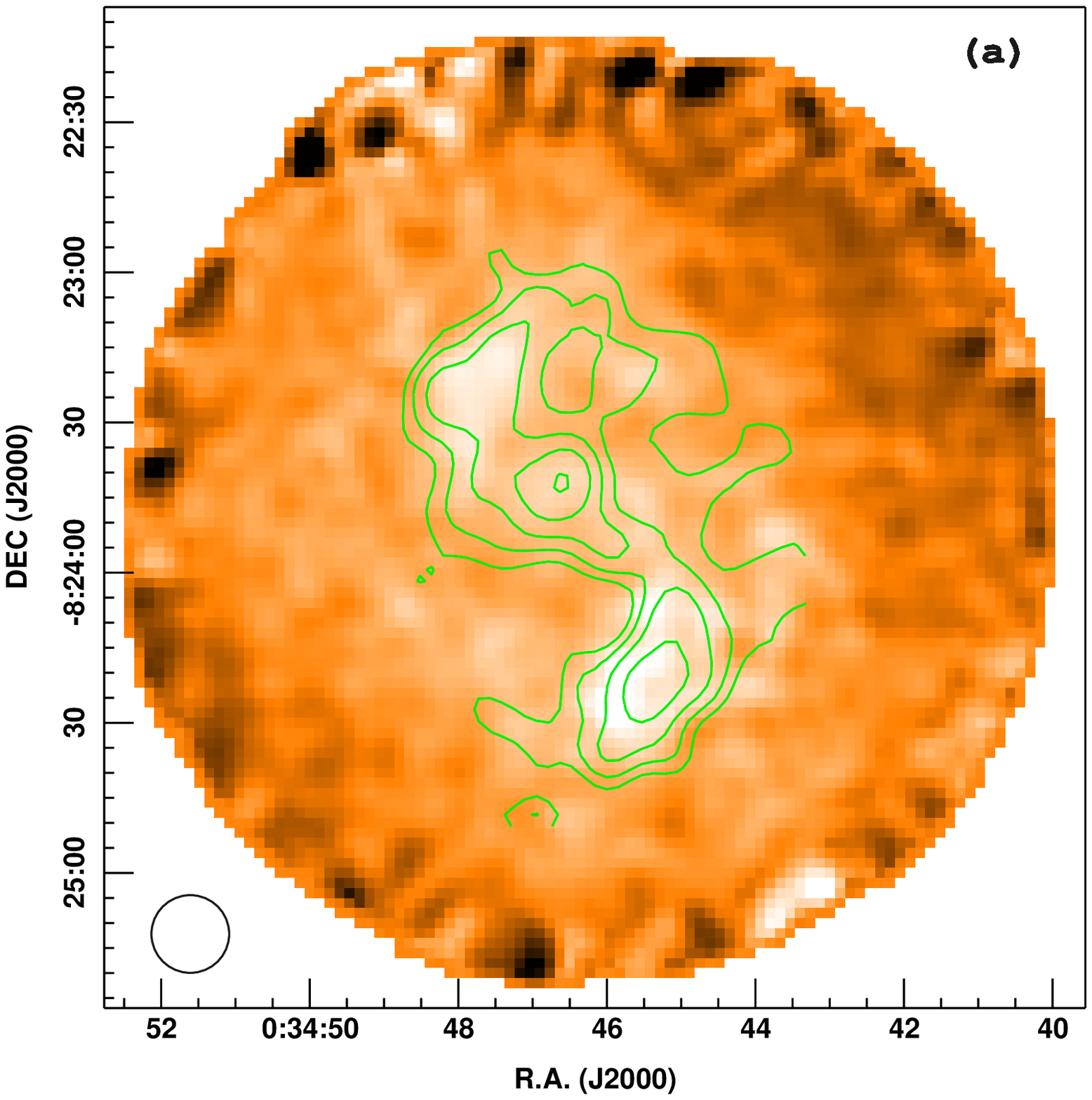}
\includegraphics[angle=0,scale=.4]{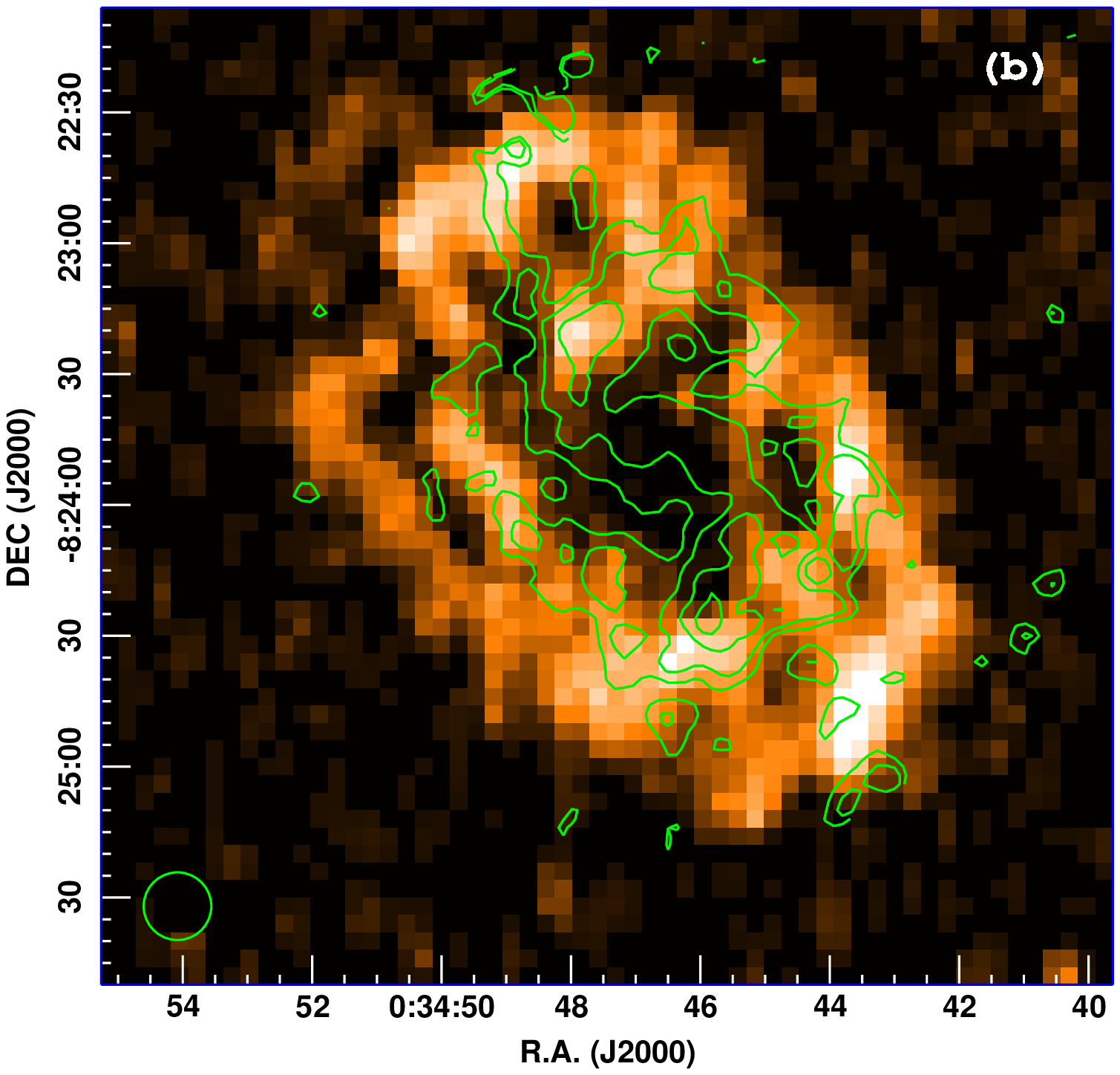}
\includegraphics[angle=0,scale=.4]{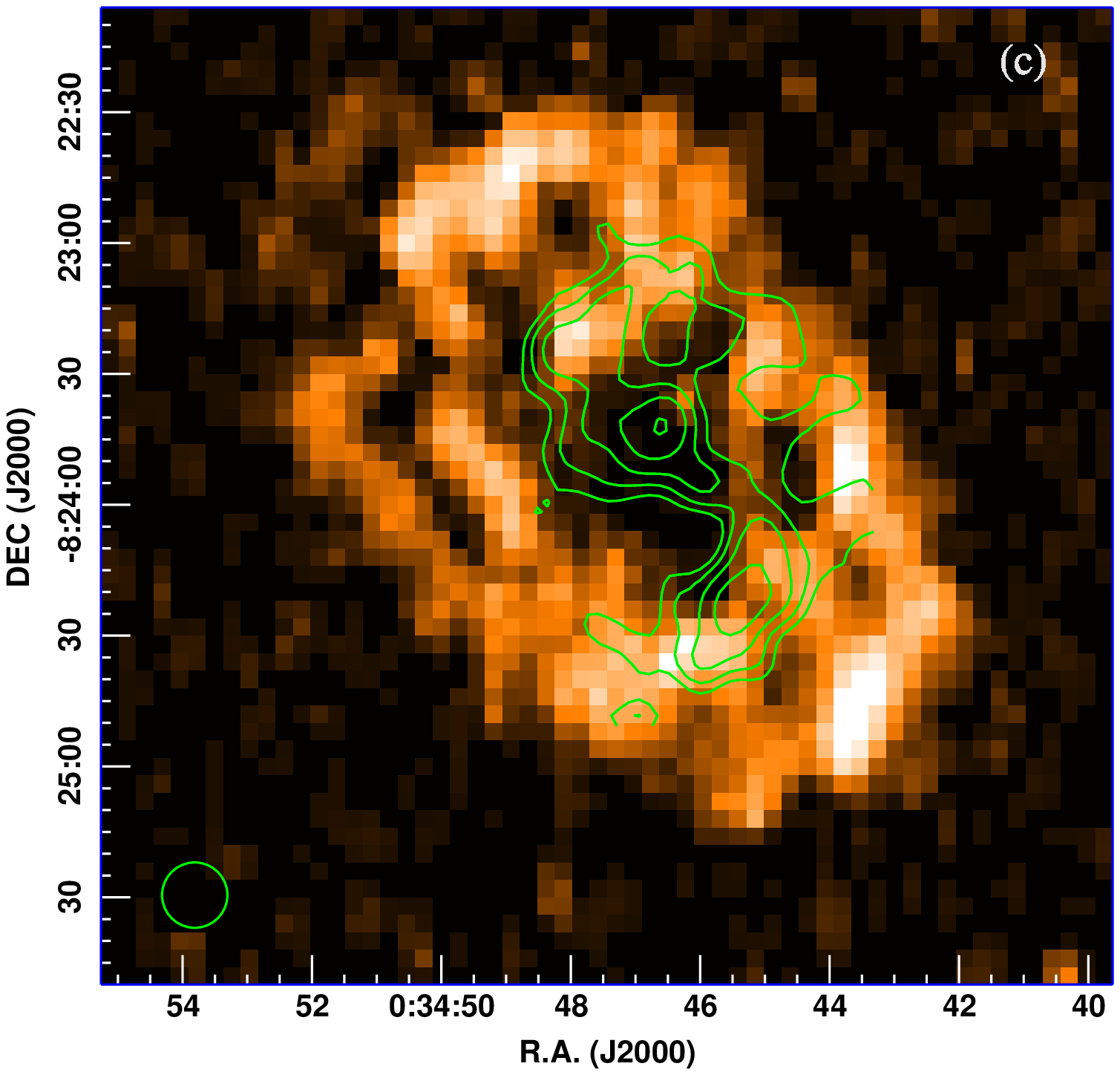}
\figcaption{
Morphological comparison between the different ISM tracers (\ion{H}{1},
$850 \mu m$ and $^{12}$CO(3--2)) for NGC\,157. 
In Figure 8a, the $^{12}$CO(3--2) contours are overlaid on top
of the SCUBA 850 \um map. The contour levels are 
4, 5.7, 8, 11.4 and 16 \kkms.
In Figure 8b, the 850 \um contours are overlaid on top of the 
the \ion{H}{1} map. The contour levels are 17, 23, 34 and 46 
mJy/beam.
Figure 8c presents the $^{12}$CO(3--2) contours on top of the
\ion{H}{1} map. The contour levels are as in Figure 8a.
In all maps the beam size ($15''$) is indicated on the bottom-left corner.
\label{fig-8}}
\end{figure}

\newpage
%Figure 9
\begin{figure}
\includegraphics[angle=0,scale=.8]{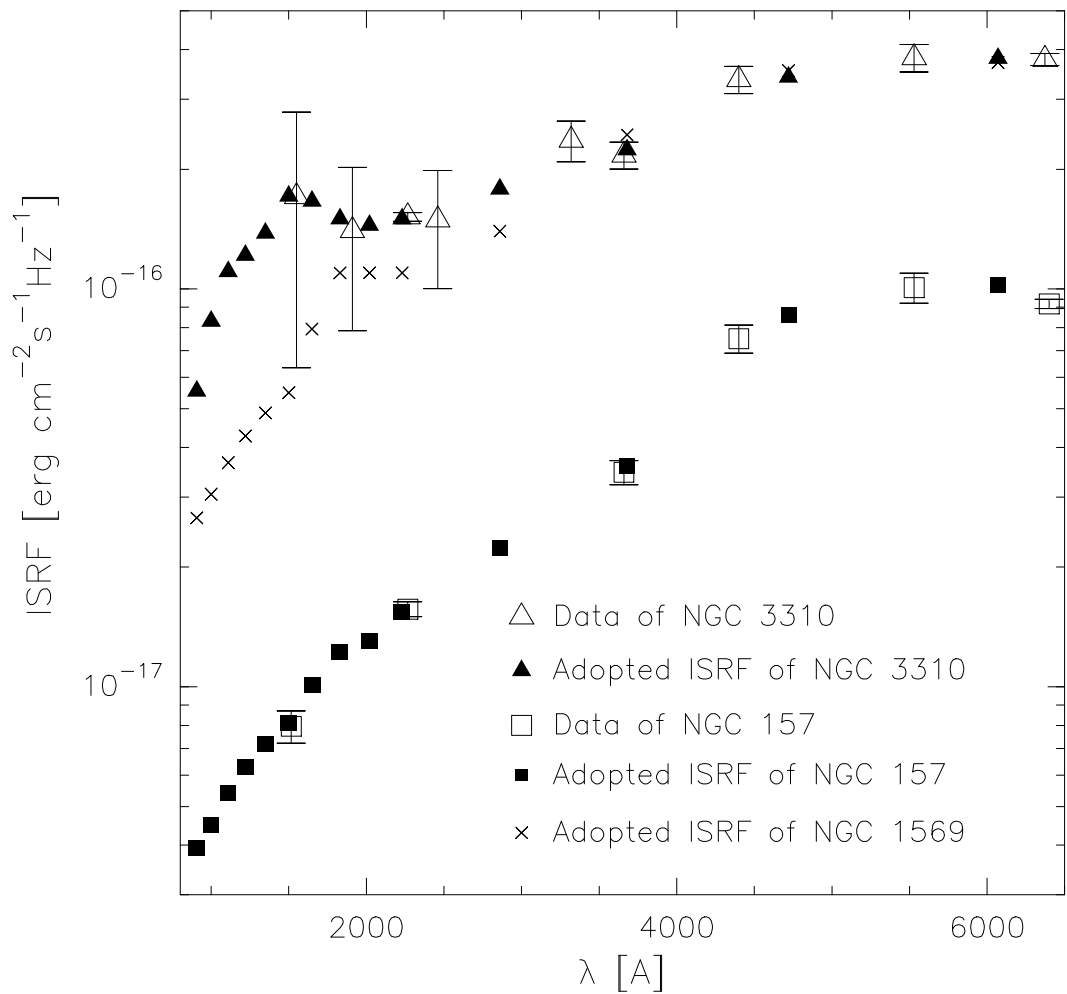}
\figcaption{
The intensity of the interstellar radiation field of NGC\,3310 
and NGC\,157 compared to that of the low-metallicity dwarf
galaxy NGC\,1569 (see Appendix A for details). 
The open symbols (triangle
and square) are derived from the observed data listed in Table 10,  
assuming
the emitting region as described in Appendix A.
The filled symbols give the interpolation to the observed data points
used as input in the DBP90 model. The crosses give the interpolated
ISRF of the dwarf starburst galaxy NGC\,1569 for comparison.
\label{fig-9}}
\end{figure}

%\newpage
%\input{./tables_astroph.tex}

\end{document}

%%%%%%%%%%%%%%%%%%%%%%%%%%% TABLE 1 %%%%%%%%%%%%%%%%%%%%%
\begin{table}
\begin{minipage}{8.7truecm}
\begin{center}
\caption{Calibration data of the sub-mm SCUBA observations.}
\begin{tabular}{cccc} \hline
Wavelength & $\theta_{\rm HPBW}$ & $G, \delta G / G$ \footnote{The flux gain
in units of Jy  beam$^{-1}$ V$^{-1}$ and its fractional uncertainty
estimated from an extensive set of beam maps on CRL\,618.}& $\sigma_{\rm SKY}$
\footnote{The sky $\sigma$ in the final images.}\\
($\mu$m)   & (arcsec)        & ($\%$) & (mJy beam$^{-1}$)  \\
\hline
850  & 15.28 & 221, 16 & 4.2\\
450  & 9.28  & 332, 32 & 29.0\\
\hline
\end{tabular}
\end{center}
\end{minipage}
\end{table}
%%%%%%%%%%%%%%%%%%%%%%%%%%%%%%%%%%%%%%%%%%%%%%%%%%%%%%%%%%%%%%%%%%%%%%%%%

%=======Table 2=======
{\small
\begin{table}
\begin{center} 
\caption{$^{12}$CO(1--0) and $^{12}$CO(3--2) integrated intensities in NGC\,3310.}
              
\vskip 2.ex
\tabcolsep 3pt
\begin{tabular}{lcccccc}
\hline
\hline
{ X } &
{ Y  } &
{$I_{10}$ $^a$} &
{$\sigma_{10} $ } &
{$I_{32}$ $^a$} &
{$\sigma_{32} $ } &
{$r_{31}$  $^b$} 
\\
{$''$ } &
{$''$ } &
{\rm K km s$^{-1}$} &
{\% } &
{\rm K km s$^{-1}$} &
{\% } &
{ }
\\
\hline
%------------------------------------------------------------
%x	y	Ico10 		sigma1 I32Tmb  sigma32  r31_15
%--------------------------------------------------
21  &	7   &	$<$5   &	--    & 3.8   &	30  &$>$0.76	      \\
21  &	0   &	$<$5   &	--    & 3.9   &	25  &$>$0.78	      \\
14  &	21  &	4.3  &	47   & 3.5   &	19  &	0.81  \\
14  &	14  &	5.2  &	38   & 5.32  &	14  &	1.02  \\
14  &	7   &	5.6  &	25   & 4.6   &	35  &	0.82  \\
14  &	0   &	8.2  &	20   & 11.29 &	16  &	1.38  \\
14  &	-7  &	9    &	19   & 5.48  &	12  &	0.61  \\	
14  &	-14 &	$<$5   &	--    & 4.3   &	29  &	$>$0.86\\
7   &	21  &	6.5  &	32   & 5.60   &	19  &	0.86  \\	
7   &	14  &	8    &	21   & 9.73  &	12  &	1.22  \\	
7   &	7   &	11.3 &	16   & 11.45 &	11  &	1.01  \\	
7   &	0   &	10.7&	15   & 11.13 &	14  &	1.04  \\	
7   &	-7  &	5.4  &	49   & 8.5   &	21  &	1.57  \\
7   &	-14 &	$<$5   &	--    & 5.81  &	16  &	$>$1.16  \\	
0   &	21  &	9.9  &	19   & 8.39  &	14  &	0.85  \\
0   &	14  &	4.8  &	32   & 7.16  &	13  &	1.49  \\
0   &	7   &	14.5 &	15   & 14.35 &	12  &	0.99  \\
0   &	0   &	9.3  &	15   & 10.68 &	11  &	1.15  \\
0   &	-7  &	6.8  &	17   & 8.65  &	11  &	1.27  \\
0   &	-14 &	4.8 &	39   & 5.53  &	30  &	1.15  \\
0   &	-21 &	4.9  &	42   & 4.52  &	33  &	0.92  \\
-7  &	21  &	$<$5   &	--    & 4.1   &	20  &	$>$0.82  \\
-7  &	14  &	$<$5   &	--    & 7.5   &	16  &	$>$1.50  \\
-7  &	7   &	$<$6   &	--    & 9.79  &	11  &	$>$1.63  \\
-7  &	0   &	$<$6   &	--   & 7.76  &	15  &	$>$1.29  \\
-7  &	-7  &	$<$6   &	--   & 11.7  &	15  &	$>$1.95 \\
-7  &	-14 &	5.0  &	27   & 6.77  &	12  &	1.35  \\
-7  &	-21 &	$<$5   &	--    & 4.0   &	20  &	$>$0.80   \\	
-14 &	14  &	$<$5   &	--    & 1.74  &	22  &	$>$0.35  \\
-14 &	7   &	$<$5   &	--    & 5.1   &	25  &	$>$1.02  \\
-14 &	0   &	$<$5   &	--    & 4.05  &	15  &	$>$0.81  \\
-14 &	-7  &	4.3  &	47   & 5.89  &	12  &	1.37  \\
-14 &	-14 &	$<$5   &	--    & 7.3   &	15  &	$>$1.46  \\
%--------------------------------------------------
\hline
\end{tabular}
\end{center}
\noindent 
$ ^{\rm  a}$ Both the $I_{10}$ and $I_{32}$ intensities are in main beam temperature. Non-detections for $^{12}$CO(1--0) have an upper limit of 3 sigma.
\\
$ ^{\rm  b}$ The $^{12}$CO(3--2) line intensity has been converted 
from $14''$ to $15''$ to match the $^{12}$CO(3--2) HPBW. 
\\
%(This conversion did not make much difference.)
%The map centre is J2000 10 38 45.9  dec=53 30 11.7
%or B1950 10 35 40.1 dec = 53 45 49
\end{table}
\newpage

\begin{table}
\caption{$^{12}$CO(1--0) and $^{12}$CO(3--2) integrated intensities in NGC\,157.}
\end{table}
%================= Table 3==============
{\small
%\begin{table}
\begin{center} 
              
\vskip 2.ex
\tabcolsep 3pt
\begin{tabular}{lcccccc}
\hline
\hline
{ X } &
{ Y  } &
{$I_{10}$ $^a$} &
{$\sigma_{10} $ } &
{$I_{32}$ $^a$} &
{$\sigma_{32} $ } &
{$r_{31}$  $^b$} 
\\
{$''$ } &
{$''$ } &
{\rm K km s$^{-1}$} &
{\% } &
{\rm K km s$^{-1}$} &
{\% } &
{ }
\\
\hline
%-------------------------------------------------------
%x	y	Ico10	Sigma Ico32	Sigma32	r31_15	
%-------------------------------------------------------
14  &	14  &	19.9 &	15 &	7.29 &	13  &	0.37   \\
14  &	7   &	9.3  &	24 &	8.79 &	16  &	0.95   \\
14  &	0   &	11.0 &	17 &	6.52 &	15  &	0.59   \\
14  &	-7  &	6.5  &	32 &	5.24 &	15  &	0.81   \\
14  &	-14 &	9.0  &	19 &	$<$1.8 &	--  &	$<$0.2   \\
7   &	56  &	4.5  &	41 &	$<$1.8 &	--  &	$<$0.4   \\
7   &	49  &	17.2 &	13 &	2.51 &	25  &	0.15   \\
7   &	42  &	12.9 &	22 &	4.74 &	16  &	0.37   \\
7   &	35  &	15.5 &	15 &	5.83 &	15  &	0.38   \\
7   &	28  &	15.5 &	15 &	5.0  &	17  &	0.32   \\
7   &	21  &	6.7  &	24 &	5.31 &	19  &	0.79   \\
7   &	14  &	10.0 &	22 &	6.23 &	15  &	0.62   \\
7   &	7   &	7.9  &	21 &	9.36 &	15  &	1.18   \\
7   &	0   &	13.4 &	15 &	10.99&	11  &	0.82   \\
7   &	-7  &	13.5 &	15 &	5.40 &	16  &	0.40   \\
7   &	-14 &	7.0  &	20 &	$<$1.8 &--  &	$<$0.26   \\
7   &	-21 &	$<$5.0 &-- &	$<$2.0 &--  &	--    \\
7   &	-28 &	$<$6.0 &-- &	$<$1.8 &--  &	--     \\
7   &	-35 &	11.4 &	16 &	$<$1.8 &--  &	$<$0.16   \\
7   &	-42 &	6.5  &	32 &	3.70 &	16  &	0.57   \\
7   &	-49 &	9.5  &	19 &	4.12 &	16  &	0.43   \\
7   &	-56 &	5.8  &	23 &	3.07 &	19  &	0.53   \\
7   &	-63 &	$<$6.0 &	-- &	6.08 &	23  &	$>$1.01  \\
0   &	63  &	6.2  &	26 &	$<$1.8 &	--  &	$<$0.29   \\
0   &	56  &	4.2  &	33 &	2.80 &	22  &	0.67   \\
0   &	49  &	4.8  &	33 &	3.64 &	19  &	0.76   \\
0   &	42  &	8.7  &	23 &	4.74 &	17  &	0.54   \\
0   &	35  &	11.5 &	16 &	5.18 &	15  &	0.45   \\
0   &	28  &	5.2  &	31 &	4.43 &	15  &	0.85   \\
0   &	21  &	6.3  &	26 &	5.11 &	16  &	0.81   \\
0   &	14  &	7.8  &	27 &	7.38 &	14  &	0.94   \\
0   &	7   &	11.8 &	19 &	14.3 &	11  &	1.21   \\
0   &	0   &	7.8  &	30 &	10.25&	12  &	1.31   \\
0   &	-7  &	10.8 &	17 &	6.12 &	14  &	0.57   \\
\hline
\end{tabular}

\clearpage
\begin{tabular}{lcccccc}
\hline
\hline
{ X } &
{ Y  } &
{$I_{10}$ $^a$} &
{$\sigma_{10} $ } &
{$I_{32}$ $^a$} &
{$\sigma_{32} $ } &
{$r_{31}$  $^b$} 
\\
{$''$ } &
{$''$ } &
{\rm K km s$^{-1}$} &
{\% } &
{\rm K km s$^{-1}$} &
{\% } &
{ }
\\
\hline
%-------------------------------------------------------
%x	y	Ico10	Sigma Ico32	Sigma32	r31_15	
%-------------------------------------------------------
0   &	-14 &	14.9 &	17 &	2.46 &	24  &	0.17   \\
0   &	-21 &	6.0  &	19 &	3.42 &	15  &	0.57   \\
0   &	-28 &	9.8  &	19 &	$<$1.8 &	--  &	$<$0.18  \\
0   &	-35 &	6.3  &	19 &	4.32 &	16  &	0.69   \\
0   &	-42 &	6.6  &	29 &	4.19 &	18  &	0.63   \\
0   &	-49 &	11.8 &	16 &	3.49 &	19  &	0.30   \\
0   &	-56 &	4.7  &	33 &	$<$2.0 &	--  &	$<$0.42   \\
0   &	-63 &	$<$5.0 &	-- &	$<$2.0 &	--  &	 --    \\
-7  &	56  &	3.6  &	43 &	2.25 &	28  &	0.63   \\
-7  &	49  &	$<$5.0 &	-- &	3.21 &	19  &	$>$0.64  \\
-7  &	42  &	6.2  &	26 &	6.61 &	13  &	1.07   \\
-7  &	35  &	7.5  &	17 &	5.36 &	16  &	0.71   \\
-7  &	28  &	12.8 &	15 &	5.74 &	14  &	0.45   \\
-7  &	21  &	$<$5.0 &	-- &	5.31 &	17  &	$>$1.06  \\
-7  &	14  &	7.1  &	20 &	4.42 &	19  &	0.62   \\
-7  &	7   &	7.5  &	16 &	6.13 &	15  &	0.82   \\
-7  &	0   &	5.5  &	24 &	5.38 &	20  &	0.98   \\
-7  &	-7  &	$<$6.0 &	-- &	7.57 &	12  &	$>$1.26  \\
-7  &	-14 &	10.8 &	15 &	$<$2.1 &	--  &	$<$0.19   \\
-7  &	-21 &	7.0  &	29 &	$<$1.7 &	--  &	$<$0.24   \\
-7  &	-28 &	9.8  &	21 &	2.48 &	28  &	0.25   \\
-7  &	-35 &	13.8 &	15 &	6.44 &	14  &	0.47   \\
-7  &	-42 &	35.1 &	15 &	6.74 &	14  &	0.19   \\
-7  &	-49 &	7.0  &	26 &	8.38 &	18  &	1.20   \\
-7  &	-56 &	5.4  &	21 &	3.69 &	19  &	0.68   \\
-14 &	14  &	5.8  &	28 &	2.81 &	24  &	0.48   \\
-14 &	7   &	15.7 &	15 &	5.93 &	13  &	0.38   \\
-14 &	0   &	7.7  &	20 &	4.55 &	19  &	0.59   \\
-14 &	-7  &	15.0 &	17 &	6.14 &	14  &	0.41   \\
-14 &	-14 &	14.8 &	17 &	5.79 &	15  &	0.39   \\
%-------------------------------------------------------------------------
\hline
\end{tabular}
\end{center}
\noindent
$ ^{\rm  a}$ Both the $I_{10}$ and $I_{32}$ intensities are in main beam temperature.  For non-detections, upper limits of 3 sigma are listed.
\\
$ ^{\rm  b}$ The $^{12}$CO(3--2) line intensity has been converted 
from $14''$ to $15''$ to match the $^{12}$CO(3--2) HPBW. 
\\
%The map centre is J2000 00 34 46.5  dec=-8 40 23 48
%or B1950 00 32 14.18 dec = -8 40 19.4	    
%end{table}
\newpage

%=============Table 4 ==========
{\small
\begin{table}
\begin{minipage}{10truecm}
\begin{center} 
\caption{Integrated intensity ratios at the central $21''$ region.}
\vskip 2.ex
\tabcolsep 3pt
\begin{tabular}{lcccccc}
\hline
\hline
{ Source } &
{$I_{10}$ } &
{$I_{21}$ } &
{$I_{32}$ } &
{$ r_{21}$ } &
{$ r_{31}$ ~ $^a$} &
{$ r_{31}$ ~ $^b$}  
\\
{   } &
{\rm K km s$^{-1}$} &
{\rm K km s$^{-1}$} &
{\rm K km s$^{-1}$} &
{ } &
{ } &
{ }
\\
\hline
%-------------------------------------------------------
%obj        I10     I21    I32    r21   r31_21''  r31_15''
%         (Tmb)     K km/s
%---------------------------------------------------------
NGC\,157     &  9.7  &  9.7 &    8.34 &  1.0  &   0.86 &    1.28 \\
NGC\,3310    &  6.62 &  9.8 &    9.88 &  1.48 &   1.49 &    1.17 \\
%N157SW   &       &      &         &       &        &         \\
%(-14,-7) &       &      &         &  0.43 &   0.46 &    0.4  \\
%(assuming the IRAM data is for this point)
%---------------------------------------------------------
\hline
\end{tabular}
\end{center}
\noindent 
$ ^{\rm  a}$ The $r_{31}$ ratio at a $21''$ resolution.
\\
$ ^{\rm  b}$ The $r_{31}$ ratio at a $15''$ resolution.
\\
%For N157 R21 > 7 we can use ~10.

\end{minipage}
\end{table}

%%%%%%%%%%%%%%%%%%%%%%%%%%% TABLE 5 %%%%%%%%%%%%%%%%%%%%%
\begin{table}
\begin{minipage}{13truecm}
\begin{center} 
\caption{LVG Analysis Results.}
\vskip 2.ex
\tabcolsep 3pt
\begin{tabular}{lccccccc}
\hline
\hline
{Source} &
{\nh2~$^a$} &
{\Tk~$^a$} &
{$Z_{\rm CO}/(dV/dr)$~$^a$} &
$\tau_{10}~^b$  &
$X$~$^c$             &
$\rm K_{vir}~^d$ &
{$\chi^2 ~^e  $} 
\\
{} &
{$10^3$ \cm3} &
{K} &
{$ 10^{-6}$ pc (km/s)$^{-1}$} &
{ } &
{ $10^{19}$ } &
{} &
{}
\\ 
\hline  
NGC\,3310    &  7.8       &  50     &    2.7    &    0.3     &  13.0     &10.1   &   0.9\\ 
         &  5.0--8.0  &  40--60 &1.0--2.7   &  0.18--0.75  & 6.3--13.0 &10--15   &   $<$2 \\
 
NGC\,157C  &  3.5         &  30    &   2.7     &    0.7        & 4.6   &6.1   &   0.6\\ 
         &  2.0-- 5.0 &  30--40 & 2.7-- 7.0 &  1.3--1.6     &4.6--6.1 &6--13   &   $<$2 \\ 

NGC\,157S  &  1.0--2.0    &  10--20 &  6--12     &  1.5--6.7    & 20--30 &3--4   &   $<$2 \\ 
\hline 
%-------------------------------------------------------------------
\end{tabular}
\end{center}
\noindent
$^a$ The LVG parameters described in \S 3.2.
\\
$^b$ The optical depth at the $^{12}$CO(1--0) transition.
\\
$^c$ The $X$ factor in the unit of 
cm$^{-2}$ ($\rm K\,km\,s^{-1}$)$^{-1}$, 
 assuming $Z_{\rm CO}= 5 \times 10^{-5}$ in NGC\,157 and 
$Z_{\rm CO}= 2 \times 10^{-5}$ in NGC\,3310.
\\
$^d$ The parameter $\rm K_{vir}$ defined in equation (4).
\\
$^e$ The value of $ \chi^2 $ associated with the fit. 

{\bf Notes}.
NGC\,157C and NGC\,157S correspond to the central region and and the southern disk 
of NGC\,157 respectively.
For each position, the first row is the best fit parameters and
the second row is the estimated range within the observed uncertainty.
For NGC\,157S the $r_{31}$ ratio is in the range 0.2 -- 0.4 and thus a range
of values is given for the parameters.
\end{minipage}
\end{table}
%%%%%%%%%%%%%%%%%%%%%%%%%%%%%%%%%%%%%%%%%%%%%%%%%%%%%%%%%%%%%%%%%%%%%%%%%

%%%-- Table 6--

\begin{table}
\begin{minipage}{13truecm}
\caption{Far-IR and sub-mm flux densities for NGC\,3310 and  NGC\,157.}
\begin{tabular}{c|cccc|cccc}
\noalign{\smallskip}
\hline
\hline
 & \multicolumn{8}{c}{Integrated flux densities} \\
  & \multicolumn{4}{c}{NGC\,3310} & \multicolumn{4}{c}{NGC\,157}\\ 
 $\lambda$     & observed  & CO & radio & dust & observed  & CO & radio & dust  \\
\ $\mu$m  & Jy  & Jy  & Jy  & Jy  & Jy  & Jy  & Jy  & Jy \\
\hline
 12  & 1.54$\pm$ 0.03  &  --      & --& 1.54$\pm$ 0.15 & 1.61$\pm$ 0.04   &  --& --&  1.61$\pm$ 0.16\\
  25 &  5.32 $\pm$0.03 &  --      & --& 5.32 $\pm$0.53 & 2.17$\pm$ 0.04 &--& --& 2.17 $\pm$ 0.22\\
  60 &  34.6 $\pm$0.05 &--        & --& 34.6 $\pm$3.46 & 17.9 $\pm$ 0.05  &--& --&   17.9  $\pm$ 1.79 \\
 100& 44.2  $\pm$0.17 & --       & --& 44.2  $\pm$4.42 & 43.4$\pm$ 0.1  &--& --& 43.4$\pm$ 4.34  \\
 450&  1.32 $\pm$0.26 & --       & 0.035 & 1.29$\pm$0.26 & 4.68$\pm$ 0.93& -- & 0.018 & 4.66 $\pm$0.93 \\
 850& 0.36 $\pm$0.07  &0.051&0.039 & 0.27 $\pm$ 0.07& 0.93$\pm$ 0.19 & 0.14 & 0.016 & 0.77$\pm$0.19    \\
\hline
\label{dust_sed}
\end{tabular}

% CO contamination: 14% for N157, 15% for N3310
{\bf Notes.}
 The flux densities  at 12, 25, 60 and 100 $\mu$m are from IRAS (Soifer et al. 2003). The errors
 for the observed flux densities represent 1$\sigma$ uncertainties of the data,
 whereas we adopt a 10\% error as a more realistic value for the dust flux density
 at these wavelengths.
 The fluxes at 450 $\mu$m and 850 $\mu$m are derived by us by the integration
 over the total extent of the  corresponding maps from the JCMT archive.
 In order to derive the flux density corresponding to the pure dust emission (column ``dust") we have to
 correct for the contamination of the $^{12}$CO(3--2) at 850 $\mu$m (column ``CO") 
and the radio contamination (column ``radio") at sub-mm wavelengths 
(see \S 3.1.1). The corresponding values are listed.
\end{minipage}
 \end{table} 

%%% 

%==========--table 7 ----

{\small
\begin{table}
\begin{center} 
\caption{ SED fitting result from the 2T model.}

\vskip 2.ex
\tabcolsep 3pt
\begin{tabular}{lccccc}
\hline
\hline
{Galaxy } &
{$T_{\rm W}$ } &
{$T_{\rm C}$ } &
{$M_{\rm w, dust}$ }  &
{$M_{\rm c, dust}$ }  &
{$M_{\rm  dust}$ } \\
{ } &
{\rm K } &
{\rm K } &
{$10^8 \ms$} &
{$10^8 \ms$}  &
{$10^8 \ms$} \\
\hline
NGC\,3310   &   34.4    & 5.8 &   0.05  & 2.19  & 2.24 \\ 
NGC\,157    &   27.9    & 10.6  &   0.18 & 2.34 & 2.52 \\
\hline
\end{tabular}
\end{center}

\noindent

\end{table}

%==========--table 8 ----

{\small
\begin{table}
\begin{center} 
\caption{ SED fitting result from the DBP90 model.}

\vskip 2.ex
\tabcolsep 3pt
\begin{tabular}{lcccc}
\hline
\hline
{Galaxy } &
{$T_{\rm dust}$ } &
{$M_{\rm LG}$ } &
{$M_{\rm VSG}$ } &
{$M_{\rm PAH}$ } \\
{ } &
{\rm K } &
{$ 10^6 \ms$} &
{$ 10^6 \ms$} &
{$ 10^6 \ms$} \\
\hline
%-----------------------------------------------------------
%           Tk    M_LG     M_VSG    M_PAH    (Tw,Tc,Mw,Mc)
%           K     10^6     10^6     10^6Msun
%-----------------------------------------------------------
%N3310      32    2.1      0.5       0.1   (33, 6,  7.2, 1.5)
%N157       22.5  54       4.5       4.1   (28, 11, 4.1,7.7)
%-------------------------------------------------------------
NGC\,3310   &   32.0    & 2.1 &     0.5 &      0.1  \\ 
NGC\,157    &   22.5  & 54  &     4.5 &      4.1  \\
\hline
\end{tabular}
\end{center}

\noindent

\end{table}

}

%==========--table 9 ----

{\small
\begin{table}
\begin{center} 
\caption{ Mass of ISM components and the gas-to-dust ratios }
              
\vskip 2.ex
\tabcolsep 3pt
\begin{tabular}{lccccccccc}
\hline
\hline
{Position$^a$} &
{$S_{\rm HI}$ } &
{$S_{\rm ^{12}CO(1-0)}$ $^b$} &
{$M_{\rm HI} $ } &
{$M(\rm H_2)$ $^c$} &
{$X ^d$} &
{$S_{850}$  $^b$} &
{$M_{\rm dust}$  $^e$} &
{$T_{\rm dust} $ $^e$} &
{$\frac{M_{\rm H_2}+M_{\rm HI}}{M_{\rm dust}}$ } 
\\
{ } &
{\rm Jy km s$^{-1}$} &
{\rm Jy km s$^{-1}$} &
{$ 10^8 \ms$} &
{$ 10^8 \ms$} &
{$ 10^{19} $} &
{\rm Jy } &
{$ 10^{7} \ms$} &
{ K }   &
{ }
\\
%------------------------------------------------------------------
%	S(HI)	S(CO)	M(HI)	M(H2)	X	S850	Mdust Td GD
%N157	Jy km/s Jy km/s  10^8	10^8	10^19   Jy	10^7  K  ratio
%-------------------------------------------------------------------
\hline
NGC\,157   &	&	&	&     &		&     &	     &	 &\\
Centre  &1.40	&71.3	&4.1	&5.16&	11	&0.091&	1.15  & 23& 79\\ 
SW	&2.28	&128.	&6.6	&16.9&	20	&0.125&	3.1  & 15& 76\\
NE     &2.03	&97.9	&5.9	&12.9&	20	&0.106&	2.6  & 15& 72\\	
$21''$  &0.11 	&45.6   &0.33   &1.68&  5.6     &0.029& 0.37 & 23& 54\\ 
Total   &78	&500	&225	&65.7 & 20	&0.8  & 23.4  & 10.6 & 124\\
%-----------------------------------------------------------------------------
\hline	    	 	      	     	 	 	    
NGC\,3310   &	&	&	&     &		&     &	     &	 &\\

$20''$  &0.83   & 28.2	&0.92   & 0.89& 13     &0.067& 0.058& VSG&  312\\
$60''$  & 8.2   & 101  & 8.8   &  3.2 & 13      &0.29  & 0.26 & VSG  & 461\\
Total   &69	&140	&73.9	& 4.4 & 13	& 0.31 & $> 0.27$ $^f$&VSG & $<$2900\\ 
%---------------------------------------------------------------------
\hline
\end{tabular}
\end{center}
\noindent
$ ^{\rm  a}$  The first three positions of NGC\,157 are in a  $42'' \times 27''$ region
               (resolution of the VLA HI data with nature weighting) 
               at the center and two 
               spiral arms $40''$ offset from the center.  
               Line 4 is for the central $21''$ region.
               In NGC\,3310 the regions are the central $20''$ and $60''$.
	       The last line for each galaxy is the total flux of
	       the whole galaxy. Note that the HI maps cover a much larger 
               area than that of CO and SCUBA 850 \um, thus the GD mass ratio
	       is an upper limit.
%$ ^{\rm  b}$  The HI flux in N157 are from the VLA map 
%               with beam size $42'' \times 27''$. In N3310 it is from 
%	       Kregel \& Sancisi (2001) at $20''$ and $60''$ resolution.
\\
$ ^{\rm  b}$  The $^{12}$CO(1--0) and SCUBA 850 \um fluxes have been convolved to match the resolution of the HI data. The  850 \um fluxes have been corrected for
$^{12}$CO(3--2) contamination, but not for thermal radio emission because its spatial
distribution is uncertain.
\\
$ ^{\rm  c}$  The molecular gas mass derived using the $X$ factor in column 6.
\\
$ ^{\rm  d}$  Our best estimate of the $X$ factor in the corresponding region, 
in the unit of {\rm\,cm$^{-2}$\,{[K km\,s$^{-1}$]}$^{-1}$}.
\\
$ ^{\rm  e}$  The dust mass derived based on our best estimate of 
$T_{\rm dust}$ with $\beta=2$ 
(NGC\,157) or an enhanced VSG model (NGC\,3310) in the corresponding region (see text for details).
\\
$ ^{\rm  f}$  The 850 \um flux here is mostly in the $60''$ region with SCUBA detection, but the HI extended to several arcminutes thus the  $S_{850}$ is
a lower limit and the GD mass ratio is an upper limit.
\\

%\vspace*{0.3cm}
\end{table}
}

%%% Table 10 - with the radiation field used

\begin{table}
\caption{Fluxes used for the calculation of the interstellar radiation field}
\begin{tabular}{ccc||ccc}
\noalign{\smallskip}
\hline
\hline
\multicolumn{3}{c}{NGC\,3310} & \multicolumn{3}{c}{NGC\,157} \\
$\lambda$ & Flux density & Ref & $\lambda$ & Flux  density & Ref. \\
\  [\AA ] & [mJy] & & [\AA ] & [mJy] &  \\
\hline
 1550  &   90$\pm$ 56  & (1) &1516 & 26$\pm$2.4 & (2)\\
 1910   &  74 $\pm$33  &  (1) & -  & - & \\
 2267   &  80$\pm$ 2& (2) &  2267 & 52$\pm$2.3 & (2) \\
 2460  &   79 $\pm$ 26 & (1)  & - & - &\\
 3320   &  125$\pm$14  &(1)  & - & - & \\
 3660 (U) &   115 $\pm$9  & (3) & 3660 (U) & 114$\pm$8   & (3) \\
 4400 (B)  &  178$\pm$14   & (3)    &  4400 (B)  & 247$\pm$20 & (3)\\
 5530  (V) &   202 $\pm$16   & (3)   &  5530 (V) & 332$\pm$ 29  & (3) \\
 6373 (R) &  200 $\pm$ 7   & (4)    & 6400(R) & 302 $\pm$ 8 & (5)\\
\hline
\end{tabular}
\\\\
\noindent
{\bf Notes.}
 The fluxes are corrected for Galactic and internal extinction by extrapolating
 the extinction in the V-band of 0.16 mag (for NGC\,3310)
 and 0.31 (for NGC\,157)  from NED to the other wavelengths
using the Draine (2003) extinction curve.  
The references for the flux densities are:  
(1) Code \& Welch (1982), 1982,
(2) Our own value derived from data from the GALEX archive,
(3) de Vaucouleur et al. (1991) (RC3),
(4) P. A. James, et al. ( 2004) 
(5) Kassin, de Jong, \&  Pogge (2006).
\label{isrf}
\end{table}